\documentclass[fleqn,usenatbib]{mnras}

\usepackage{mathptmx}

\usepackage[T1]{fontenc}
\usepackage{ae,aecompl}
\usepackage{float}

\usepackage{graphicx}	
\usepackage{amsmath}	
\usepackage{amssymb}	
\usepackage{xspace}

\newcommand{\hmpc}{h^{-1}{\rm Mpc}}

\newcommand{\kms}{\;{\rm km}\,{\rm s}^{-1}}

\newcommand{\pygad}{{\sc Pygad}}
\newcommand{\simba}{\mbox{{\sc Simba}}\xspace}
\newcommand{\gizmo}{\mbox{{\sc Gizmo}}\xspace}

\newcommand{\fedd}{f_{\rm Edd}}

\title[X-rays from \simba Groups]{Redshift Evolution of Galaxy Group X-ray Properties  in \simba}
\author[Robson \& Dav\'e]{Dylan Robson $^{1}$, Romeel Dav\'e $^{1,2,3}$
\\
\\$^1$ University of Edinburgh, Edinburgh, UK
\\$^{2}$ University of the Western Cape, Bellville, Cape Town 7535, South Africa
\\$^{3}$ South African Astronomical Observatories, Observatory, Cape Town 7925, South Africa
}
\date{Accepted XXX. Received YYY; in original form ZZZ}

\begin{document}

\label{firstpage}
\pagerange{\pageref{firstpage}--\pageref{lastpage}}

\maketitle

\begin{abstract}
We examine the evolution of intragroup gas X-ray scaling relations for group-sized halos ($M_{500}=10^{12.3-15}M_{\odot}$) in the \simba galaxy formation simulation. X-ray luminosity $L_X$ vs $M_{500}$ shows increasing deviation from self-similarity from $z=3\to 0$, with $M_{500}<10^{13.5} M_{\odot}$ halos exhibiting a large reduction in $L_X$ and slight increase in X-ray luminosity-weighted temperature $T_X$. These shifts are driven by a strong drop in $f_{\rm gas}$ with time for these halos, and coincides with the onset of black hole jet feedback in these systems at $z\sim 1.5$ in \simba. The connection with black hole feedback is corroborated by $f_{BH}\equiv M_{BH}/M_{500}$ in $M_{500}<10^{13.5} M_{\odot}$ halos being strongly anti-correlated with $L_X$ and $f_{\rm gas}$ at $z\la 1.5$. This is further reflected in the scatter of $L_X-T_X$: halos with small $f_{BH}$ lie near self-similarity, while those with the highest $f_{BH}$ lie furthest below.  Turning off jet feedback results in mostly self-similar behaviour down to $z=0$. For the X-ray weighted metallicity $Z_X$, stellar feedback impacts the enrichment of halo gas.  Finally, halo profiles show that jet feedback flattens the electron density and entropy profiles, and introduces a core in X-ray surface brightness particularly at $M_{500}<10^{13.5} M_{\odot}$. This argues that intragroup X-ray evolution is largely driven by jet feedback removing hot gas from the cores of massive groups, and expelling gas altogether in less massive groups. 
\end{abstract}

\section{Introduction}
Clusters of galaxies are formed as peaks in the primordial density field collapse and subsequently grow through hierarchical clustering. As probes of the high-mass tail of the spectrum of mass perturbations, these clusters allow us to constrain the evolution of structure in the universe. The basic properties of these clusters are largely determined by initial conditions and the dynamically dominant dissipationless dark matter, although baryonic processes can have a non-trivial impact particularly in the group regime. Models of dark matter driven dissipationless collapse are nonetheless broadly successful in predicting relationships between cluster properties, the most common of which being the self-similar model of clusters \citep{Kaiser}. In this scenario, groups and clusters are identical when scaled by mass, which is known as strong self-similarity \citep{Bower:1997}. Under strong self-similarity, the slope of the resulting scaling relations is not expected to evolve with redshift, although there is an expected evolution in the normalisation due to the changing density of the Universe. This model provides us with a clear baseline from which to examine the impact of baryonic physics in cluster formation.

As matter collapses to form these clusters gas experiences adiabatic compression and shock heating, creating a hot intracluster medium (ICM) with temperatures $T>10^6$K, which emit X-rays as a result of thermal bremsstrahlung as well as metal emission lines. This hot gas cools and settles into hydrostatic equilibrium.  Radiative cooling from dense gas in the cores of all but the most massive systems is expected to create a cooling flow that feeds star formation and accretion onto the central super-massive black hole. However, observed star formation rates (SFRs) are orders of magnitude lower than cooling flow estimates would suggest \citep{Fabian:2002}, resulting in the cluster cooling flow problem. 

Feedback from active galactic nuclei (AGN) is often proposed as the solution to the cooling flow problem \citep{Bower:2006,Croton:2006,Somerville:2008}, by injecting large amounts of energy into the ICM to keep the gas hot. Several other candidates to solve this problem have been suggested such as magnetic conduction \citep{NarayanMedvedev:2001, Fabian:2002}, Type~Ia supernova feedback \citep{Sasaki:2001}, and dynamical heating \citep{Birnboim:2007}, but the current consensus is that none of these are able to sufficiently counteract cooling.  Meanwhile, bubbles of hot gas seemingly created by AGN jets have been observed with sufficient mechanical inflation work to counteract cooling \citep{McNamara:2007} lending credence to the idea that AGN jet feedback is responsible for the heating of the intracluster medium (ICM), which quenches massive galaxies and solves the cluster flow problem. 

Modern models of groups and clusters must therefore include the impact of AGN feedback.  Among hydrodynamic simulations, Illustris \citep{Vogelsberger:2014, Genel:2014}, Magneticum \citep{Hirschmann:2014}, Horizon-AGN \citep{Dubois:2014, Volonteri:2016, Kaviraj:2017}, EAGLE \citep{Schaye:2015}, MassiveBlack \citep{Khandai:2015}, Blue Tides \citep{Feng:2016}, Romulus \citep{Tremmel:2017}, Illustris-TNG(TNG; \citep{Springel:2018}, FABLE \citep{Henden}, and \simba \citep{Simba} all include black hole growth and the energetic output resulting from black hole accretion. A plausible connection between AGN feedback and quenching of star formation was able to be established by EAGLE, TNG, FABLE, and \simba thanks to their success in broadly reproducing the observed massive red and dead galaxy population.  

Due to the difficult nature of modelling AGN, different simulations have approached the problem in a variety of ways.  EAGLE's implementation employs a single mode of AGN feedback with a fixed efficiency, injecting energy thermally at the location of the black hole (BH) proportional to its accretion rate. This model closely resembles quasar feedback and as such may be effective in breaking self similarity but may lack the effect of jets in opposing cooling and quenching star formation.  TNG conversely employs two forms of feedback, with kinetic black hole-driven winds at low accretion rates, and thermal feedback heating gas surrounding the BH at high accretion rates. FABLE employs a modified version of this, in which quasar mode feedback at high accretion rates is stored in the BH until there is enough energy to raise the temperature of a set number of particles by a given amount. Finally, \simba employs high-velocity jets similar to TNG, but does so in a stably bipolar fashion.  The nature of the torque-limited black hole accretion model used in \simba means that BHs inherently grow along galaxy-black hole scaling relations with no need for self-regulation~\citep{Angles:2013}, allowing for a more collimated AGN feedback to enact galaxy quenching.

Observations allow us to place constraints on these varying forms of AGN feedback at low redshifts, as we explored in \citet{Robson:2020}.  X-rays provide an efficient and physically motivated method of detection that allows us to probe the potential well of a cluster through the observation of its X-ray luminosity, $L_X$. Observed properties such as $L_X$ and the X-ray luminosity-weighted temperature $T_X$ allow us to establish proxies for the cluster mass through the use of scaling relations. These scaling relations have been studied in great detail for low redshifts \citep{Arnaud, Vikhlinin, Pratt, Bohringer}.
We are then able to use the halo mass to examine the space density of clusters, and more thoroughly test theoretical models of cluster formation. 

Moving to higher redshifts, X-ray observations rapidly become more challenging owing to surface brightness dimming, making it difficult to precisely know the evolution of scaling relations with redshift \citep{Reichert:2011}. Several studies have found self-similar evolution to redshift $z \sim 1$ driven primarily by the density of the Universe increasing with redshift \citep{Vikhlinin:2002,Maughan:2006}. Other works however have found evolution departing from that predicted by self-similarity \citep{Ettori:2004, Branchesi:2007}.  This lack of consensus may arise from difficulties in managing selection bias in samples of high-redshift clusters drawn from different surveys. Future X-ray telescopes such as {\it Lynx} and {\it Athena} are expected to allow us to probe larger samples to higher redshifts.

Thus it is timely to make predictions for the expected evolution of groups and clusters within self-consistent models including galaxy formation processes. Such simulations can help us to better understand what constraints on key physical processes such as AGN feedback are possible from present data, as well as improvements to be expected from future facilities.  By further examining how the hot gas content of halos evolves over redshift we can hopefully constrain not only the physical processes driving AGN feedback in these dense environments, but also determine how AGN drive the evolution of intragroup and intracluster gas.  Particularly enabled by simulations, studying the evolution of a set of simulated halos over time may help us better understand these samples and improve our ability to deal with surveys of high-z objects. 

In this paper we examine the redshift evolution of the X-ray scaling relations of halos in the \simba suite of cosmological simulations \citep{Simba}.  We also track how populations of halos binned by mass evolve along these scaling relations, in order to examine how various galaxy formation processes, particularly AGN feedback, have a differential impact with halo mass.  Finally, we examine these trends in test simulations where we turn off individual feedback modules, in order to isolate the particular physics in \simba driving the evolution.

This paper is organised as follows. In Section 2 we discuss the \simba suite of simulations and the key tools in reproducing X-ray scaling relations. In Section 3 we discuss the evolution of X-ray scaling relations from $z=3$ to $z=2$. In Section 4 we investigate the evolution of these scaling relations utilising differing levels of feedback within simulations. And finally in Section 5 we discuss the evolution of profiles both within the fiducial run, and within the $50 \hmpc$ No-Jet run of \simba to examine the impact of jet feedback.

\section{Simulations and Analysis}

\subsection{The \simba simulation}

The \simba simulation \citep{Simba} is a cosmological hydrodynamic simulation using the \gizmo code \citep{Hopkins}. The flagship simulation evolves a random cosmological volume of (100$h^-1$ Mpc)$^3$ including $1024^3$ dark matter particles and $1024^3$ gas elements evolved from initial conditions to the present day. \simba adopts a $\Lambda CDM$ cosmology with $\Omega_{\Lambda}  = 0.7,\Omega_m  = 0.3,\Omega_b  = 0.048, h = 0.68, \sigma_8 = 0.82 $, and $n_s = 0.97$ in line with \citet{Planck_Cosmology}.

An H$_2$-based model with the H$_2$ fraction computed from the \citet{KrumholzGnedin2011} sub grid model is used to model star formation from gas in a Monte Carlo fashion. Chemical enrichment is tracked from Type II supernovae (SNe), Type Ia SNe, and Asymptotic Giant Branch (AGB) stars, tracking eleven elements (H,He,C,N,O,Ne,Mg,Si,S,Ca,Fe) during the simulation. Radiative cooling (both primordial and metal-line) and photoionisation is included using {\sc Grackle-3.1} assuming a \citet{HaartMadau} ionising background, with on-the-fly self-shielding. Star formation driven galactic winds are modelled as decoupled two-phase winds with a mass loading factor scaled to stellar mass as described by the tracking of individual particles in the FIRE zoom simulations \citep{2017MNRAS.470.4698A}.
    
Black holes are seeded at $10^4 M_{\odot}$ in galaxies that exceed stellar masses of $10^{9.5}M_{\odot}$. There are two modes for the growing of black holes: Torque-limited accretion for gas with $T <10^5$K \citep{Angles:2017a} in which angular momentum is the primary bottleneck to accretion~\citep{Hopkins:2011}; and \citet{Bondi} accretion from higher temperature gas. The galaxy stellar mass threshold is motivated by findings in the FIRE project simulations showing that stellar feedback strongly suppresses black hole growth in low mass galaxies \citep{Angles:2017c}. \simba also includes AGN feedback, which we detail in the next section owing to its central importance in this work.

\subsection{AGN feedback in \simba}

With AGN feedback proposed as a significant mechanism in quenching star formation and contributing the the evolution of groups and clusters, simulations must take care in modelling their impact.  \citep{Heckman} described two main modes of AGN feedback: Radiative mode at high Eddington ratios ($f_{\rm Edd} \equiv \dot M_{\rm BH} / \dot M_{\rm Edd}$); and jet mode at lower $f_{Edd}$. \simba includes these two modes, plus X-ray feedback from and high-energy photon pressure. 

The energy output of radiative mode AGN is dominated by electromagnetic radiation emitted by the accretion disks of central super massive black holes (SMBH). The photon pressure drives multi-phase winds up to speeds of $\ga 1000\kms$ or more, entraining surrounding materials and as a result can include molecular \citep{Sturm_2011} and warm ionised \citep{Perna:2017} gas.

Jet mode AGN on the other hand produce feedback in the form of collimated jets, moving at speeds of order $\sim 10^4\kms$ on large scales, and this dominates the energetic output as bulk kinetic energy. This mode is thought to be powered by the advection dominated accretion of hot gas prevalent at low accretion, below $~1-2\%$ \citep{BestHeckman:2012} and the spin of SMBHs.

Significant emission from the BHs in both modes also provide a photon pressure on surrounding material. Cold gas, often found around radiative mode AGN, is able to quickly absorb and subsequently radiate away this X-ray photon pressure. However in systems with low cool gas fractions, the photon pressure is able to drive a spherical outward momentum.

The way in which these modes interact with their surrounding medium can have a significant impact on the growth and evolution of the galaxies and halos in which they reside. As a result \simba attempts to implement these forms of feedback in a way that mirrors observations as closely as possible. It is not only important to consider how the feedback is implemented, but when, since we are concerned here with redshift evolution.

Radiative and jet feedback in \simba is modelled as purely kinetic and bipolar, continuous outflows, ejected parallel to the axis of angular momentum of the inner disk. Using observations of ionised gas linewidths of X-ray detected AGN \citep{Perna:2017} the outflow velocity for radiative mode is parametrised in terms of black hole mass $M_{BH}$ as:
\begin{equation}
    v_{\textrm{w,EL}} = 500 + 500(\log M_{\textrm{BH}} - 6)/3 \kms  
\end{equation}
This model is similar to the AGN feedback implemented in \gizmo \citep{Angles:2017a} however includes a variable outflow velocity.

As $f_{\rm Edd}$ drops to $<0.2$ the jet feedback begins to add an additional velocity component, with a  strength depending on the Eddington ratio:
\begin{equation}
    v_{\textrm{w,jet}} = v_{\textrm{w,EL}} + 7000 \log(0.2/f_{\rm Edd}) \kms ,
\end{equation}
with the velocity increase capped to $7000\kms$ at $f_{\rm Edd}\leq$ 0.02. Additionally jet mode will only be triggered when $M_{BH} > 10^{7.5}M_\odot$ following observations in which radio jets only occur within galaxies containing black holes with $M_{BH} \gtrsim 10^8M_{\odot}$ \citep{Barisic:2017}. This limit has implications on when jet feedback typically begins in the simulations. As black holes grow in mass with time, we also see a decreasing $f_{\rm Edd}$ \citep{Angles:2014}. This combination of $M_{BH}$ beginning to cross the limit of $10^{7.5}$, and a decreasing $f_{\rm Edd}$ we expect to see jet feedback turning on at $\approx z\sim 1.5$ for central galaxies in group-sized halos.

While radiative mode winds are ejected at the ISM temperature, gas in jets is raised to the virial temperature of the halo. Jets are typically observed to be made up of synchroton-emitting plasma, eventually thermalising an around $T_{\rm vir}$ with surrounding hot gas \citep{Fabian:2012}. To avoid significant radiative losses in the dense gas surrounding the black hole, gas wind elements undergo a short hydrodynamic and radiative cooling decoupling time of $10^{-4}t_H$ where $t_H$ is the Hubble time at launch. 

X-ray feedback operates in two forms, depending on the gas it is heating. Non-ISM gas is heated directly according to the heating flux. ISM gas however is low resolution and pressurised such that any added heat would quickly be cooled away. Thus, one half of the energy is applied as a radial outward kick with the other being added as heat. X-ray feedback is only activated at low eddington ratios ($f_{\rm Edd} < 0.02$) alongside full velocity jets. In an attempt to model the ability of gas rich galaxies to absorb and radiate X-ray energy, galaxies must also have a cold gas fraction of $f_{\rm gas} < 0.2$. This heating has a minimal effect on the galaxy mass function, however contributes to fully quenching galaxies \citep{Appleby:2020}.

Alongside the fiducial $100\hmpc$ box, we have also run a number of versions excluding various modules of the implemented feedback. Due to the computational requirements of running these simulations, these runs use a $50\hmpc$ box with $512^3$ dark matter particles and $512^3$ gas elements. This gives these runs the same resolution as the fiducial run, with $\frac{1}{8}$th the volume. While this reduces the number of high mass halos, we are still able to examine the impact of the various forms of feedback amongst group-sized halos. The various models run are: a "No-X" model (\simba-NoX) with only X-ray feedback turned off and jets are left on; a "No-Jet" model (\simba-NoJet) with jet and associated X-ray feedback turned off; a "No-AGN" model in which all AGN feedback is turned off; and finally a "No-Feedback" model in which all feedback including star formation winds is turned off. The initial conditions for all $50\hmpc$ runs are identical.

\subsection{Identifying halos and computing X-ray emission}

Halos are identified via a 3-D friends-of-friends (FoF) finder implemented within \gizmo, using a linking length that is 0.2 times the mean interparticle spacing.  Galaxies are identified in post-processing from within these halos, using a 6-D FoF finder applied to star-forming gas and stars.  The galaxies and halos are identified and catalogued using the {\sc Caesar} code \citep{Thompson:2017}.  These catalogs, along with particle snapshots, for all the runs described above are publicly available at {\tt simba.roe.ac.uk}.

The python package \pygad\ \citep{pygad} is a multi-purpose analysis toolkit for particle-based hydrodynamics simulations, and gives us tools for the general analysis of simulations. Various criteria such as specific particles, defined regions, or particles that fit within a specified parameter range (e.g. temperature or metallicity) can be used to create sub-snapshots with particles contained within individual {\sc Caesar} halos.

A module for analysing X-ray luminosity is included in \pygad, using XSPEC \citep{Arnaud} to compute X-ray spectra. XSPEC is fully described in \citet{Eisenreich_2017}. \pygad\ calculates the X-ray luminosity of selected gas particles using emission tables in XSPEC along with particle temperatures and metallicities taken from the simulation. In our analysis of \simba's X-ray properties we use the 0.5-2keV X-ray table to be consistent with the work done in \citet{Robson:2020}. \pygad\ is publicly available at {\tt https://bitbucket.org/broett/pygad}.

In the end, we identify 1379, 1051, 509, and 126 halos with $M_{\rm 500}>10^{12.3}M_\odot$ at $z=0,1,2,3$, respectively, in our flagship run.  This constitutes the main sample that we will analyse for this work.

\section{ Evolution of Scaling Relations}

We begin by examining the evolution of X-ray scaling relations in our selected galaxy groups.  Intragroup gas can be characterised by it's X-ray luminosity $L_X$, X-ray luminosity-weighted temperature $T_X$, the hot gas fraction $f_{\rm gas}$, and the X-ray luminosity-weighted metallicity $Z_X$.  To understand the physical drivers behind these relations, we first examine the evolution with respect to $M_{500}$, both by looking at the scaling relations at various redshifts, and by tracking individual halos back in time.  In the end, we make predictions for the evolution in the most common observable barometer of hot bound gas, the $L_X-T_X$ relation, and use the evolution in the physical quantities vs. $M_{500}$ to provide a context for the predicted evolution in $L_X-T_X$.

\subsection{$L_X-M_{500}$ evolution}

 \begin{figure}
     \centering
     \includegraphics[width = 0.5\textwidth, trim = 50 0 100 0, clip]{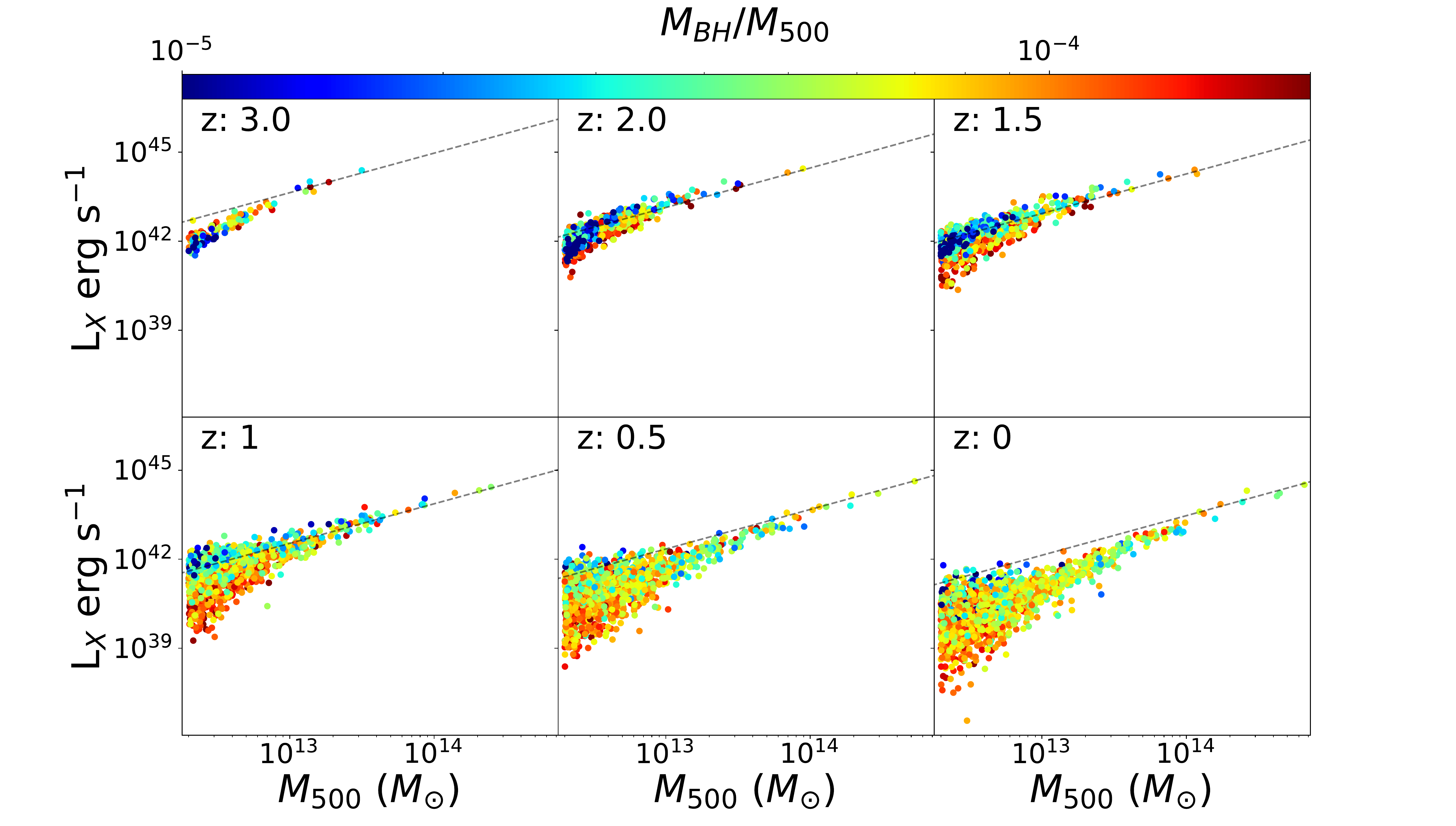}
     \caption{$L_X-M_{500}$ scaling relation from $z=3$ to $z=0$ at intervals of 0.5. Colourbar is scaled as $M_{BH}/M_{500}$. The self-similar scaling normalised to the most massive halos is shown by the dashed line. The slope of the scaling relation shows little evolution. As we move from $z=3$ to $z=0$ the most luminous halos appear to increase in mass, while the lower luminosity halos populate the lower mass end. A significant increase in scatter is also seen in halos with $M_{500} < 10^{13.5} M_{\odot}$ between $z=1.5$ and $z=1$ when jet feedback starts suggesting jet feedback may have a significant effect on these low mass halos. At a given mass, lower luminosities increasingly coincide with higher $M_{BH}$ from $z=1.5$ to $z=0$}
     
     \label{fig:M500_Lx_Scaling_Evolution}
 
 \end{figure}

\begin{figure}
    \centering

    \includegraphics[width = 0.5\textwidth]{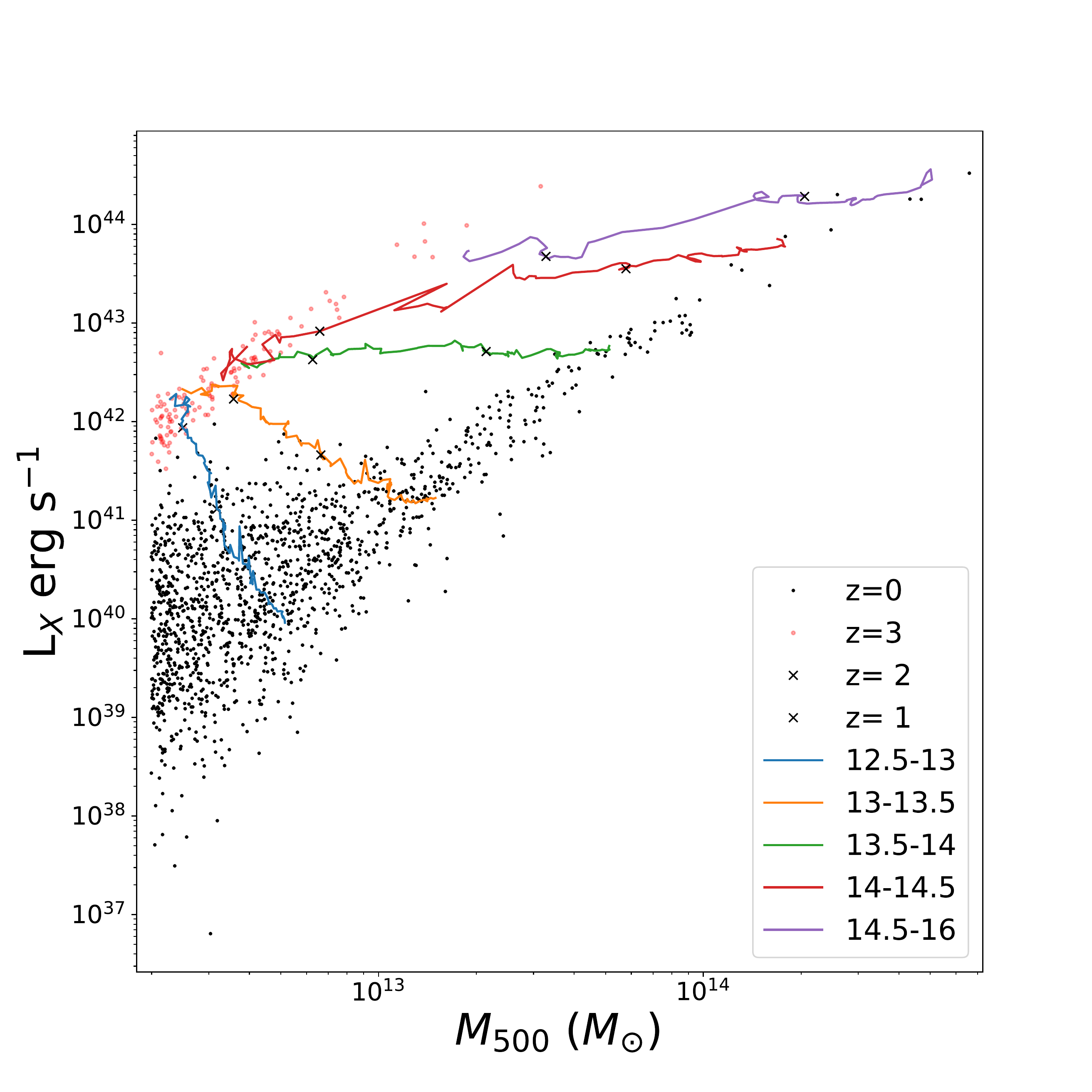}
    \caption{A plot showing the median evolution of a halo from $z=3$ to $z=0$, binned by final $M_{500}$ with blue, yellow, green, red, and finally purple representing least to most massive mass bins. X marks indicate the positions of $z=2$ and $z=1$ on each track. Light grey dots represent the scaling relation at $z=3$ while black dots represent the scaling relation at $z=0$. Halos with final masses $> 10^{13}M_{\odot}$ show little change in luminosity over an order of magnitude increase in $M_{500}$. Halos with final masses $< 10^{13}M_{\odot}$ show an order of magnitude decrease in luminosity. This indicates that the main driver behind significant evolution of $L_X$ mostly impacts the lower mass halos.}
    \label{fig:ScalingTrack_M_L}
\end{figure}

Figure \ref{fig:M500_Lx_Scaling_Evolution} shows the $L_X-M_{500}$ scaling relation for all halos with $M_{\rm halo}>10^{12.3}M_\odot$ in \simba at $z=3,2.5,2,1.5,1,0.5,0$.  Points colored by the black hole mass fraction  $f_{BH}\equiv M_{BH}/M_{500}$, where $M_{BH}$ is the black hole mass of the central galaxy.  The dashed line shows the scaling predicted from the self-similar model, normalized to the most massive halo at that epoch.

A clear $L_X-M_{500}$ is already in place at high redshifts, even though the halos are still relatively small.  As time progresses, the overall scaling relation does not change much, but there is a marked increase in the scatter, moreso at low $M_{500}$.  This starts to become particularly prominent at $z\la 1.5$.
At any given redshift, the most luminous X-ray halos at a given $M_{500}$ follow self-similarity, but there is an increasing number of halos that deviate strongly from it.

A clear trend can be seen in the $f_{BH}$ of lower mass halos. From $z=1.5\to 0$ we see the drop in $L_X$ of low mass halos correspond to an increase in $f_{BH}$.

These results can be understood by appealing to the impact of AGN jet feedback in \simba.  As noted earlier, jet feedback turns on at low $f_{\rm Edd}$.  As shown in \citet{Angles:2013}, at these mass scales, the Eddington ratio becomes comparable to the jet threshold at $z\sim 1.5$, albeit with a fair amount of scatter.  Thus we expect that black holes that have managed to grow large by these epochs will be putting out a significant fraction of their AGN feedback in high-energy jets into intragroup gas.  This circumstantially correlates well with driving a decreasing $L_X$ in systems with large $f_{BH}$ at $z\la 1.5$.  We will show more evidence for this connection with jet feedback when examining the other scaling relations.

Figure \ref{fig:ScalingTrack_M_L} provide another view on how the $z=3$ population evolves to $z=0$.  Here we examine the median evolutionary tracks of halos binned by mass.  The small points are in two groupings: The high-$L_X$ grouping (red points) is at $z=3$ while the low grouping (black points) shows the individual halos at $z=0$.  The coloured lines show median tracks of halos in various $z=0$ $M_{500}$ bins:  $M_{500}>10^{14.5}M_\odot$ (purple),  $10^{14}<M_{500}<10^{14.5}M_\odot$ (red), $10^{13.5}<M_{500}<10^{14}M_\odot$ (green), $10^{13}<M_{500}<10^{13.5}M_\odot$ (gold),
$10^{12.5}<M_{500}<10^{13}M_\odot$ (blue).  These are constructed by following individual halos back in time via their most massive dark matter progenitor, so long as the progenitor has $M_{500}>10^{12.5}M_\odot$. 

This shows that halos with $M_{500} > 10^{13.5} M_{\odot}$ (at $z=0$) show little change in $L_X$ despite an order of magnitude increase in $M_{500}$ from $z=3$ to $z=0$. In contrast, halos with a $z=0$ $M_{500} < 10^{13}$ show over an order of magnitude decrease in $L_X$, with a smaller increase in mass.  The slope of the scaling relation is maintained by low mass halos losing luminosity, along with younger halos forming at the low mass end.  In effect, the $L_X$ scaling relation spreads out in $M_{500}$ over time.

We checked that including only $z=0$ halos with progenitors all the way back to $z=3$ revealed similar trend in the median tracks.  However, if one only follows the main descendants of halos from $z=3\to 0$, this reveals that individual scaling tracks almost exclusively retain a constant luminosity as mass increases from $z=3$ to $z=0$. This indicates that the $L_X-M_{500}$ relation evolves via a combination of massive halos gaining hot gas compensated by the lowering density of the cosmos, along with low-mass halos entering into our sample above our mass cut but containing less hot gas. We will later show that this owes primarily to jet feedback removing hot gas from these lower mass halos, as found in \citet{Robson:2020}.

\subsection{$T_X-M_{500}$ Evolution}

 \begin{figure}
     \centering
     \includegraphics[width = 0.5\textwidth, trim = 50 0 100 0, clip]{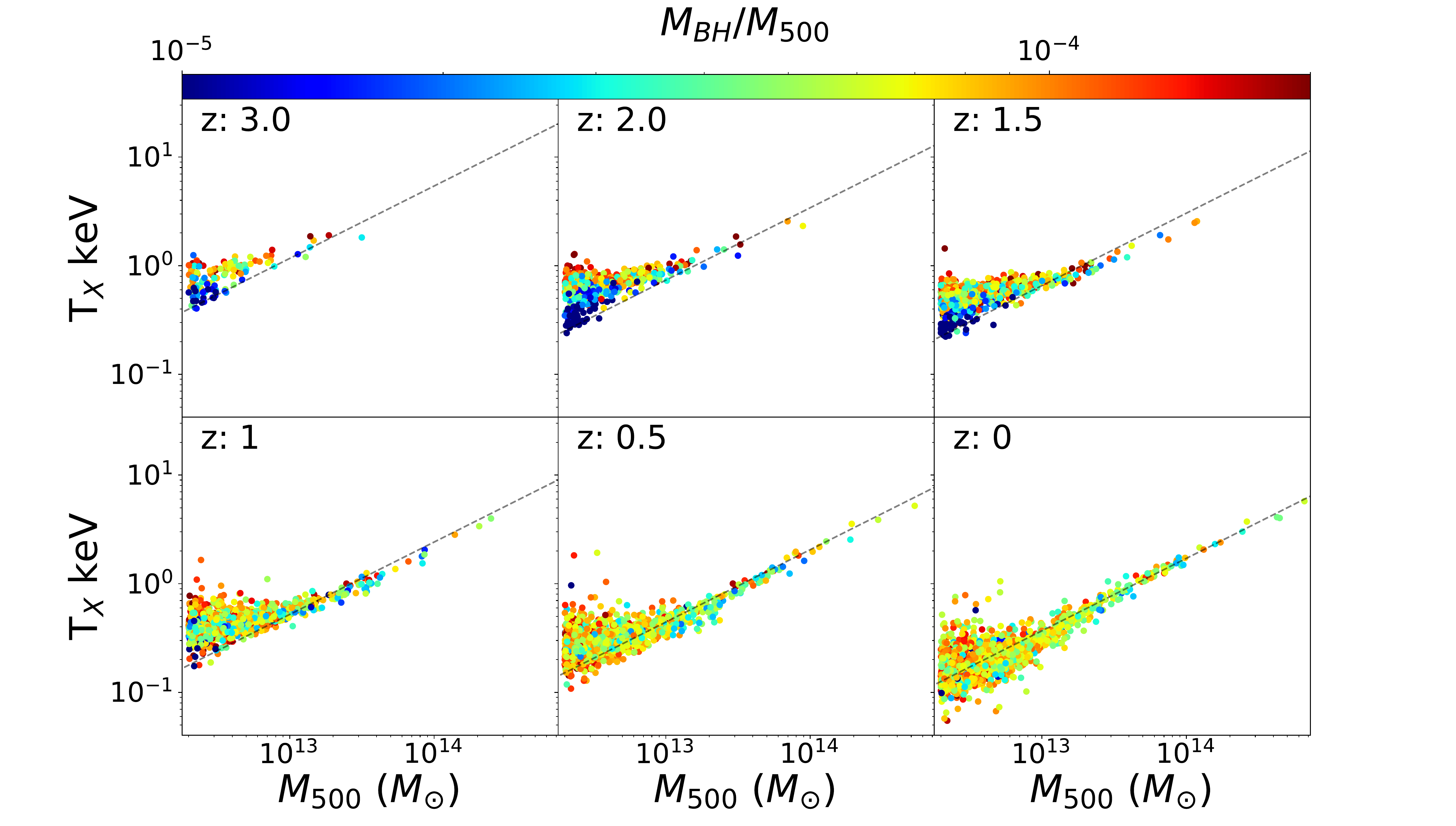}
     \caption{$T_X-M_{500}$ scaling relation from $z=3$ to $z=0$ at intervals of 0.5. The colourbar is scaled as $M_{BH}/M_{500}$. The expected self-similar scaling normalised to the most massive halos is shown by the dashed line. While the slope of the scaling relation at $M_{500} > 10^{13} M_{\odot}$ shows little evolution, a drop in $T_X$ is introduced at masses below this threshold. This drop in $T_X$ coincides with the occurrence of jet feedback indicating that jets  may be working to lower the temperature of small halos, potentially through the removal of hot gas. From $z=3$ to $z=1.5$ lower $T_X$ corresponds to a lower $M_{BH}$, while from $z=1$ to $z=0$ $M_{BH}$ increases with increasing $M_{500}$.
     }
     
     \label{fig:M500_Tx_Scaling_Evolution}
 
 \end{figure}

\begin{figure}
    \centering
    \includegraphics[width = 0.5\textwidth]{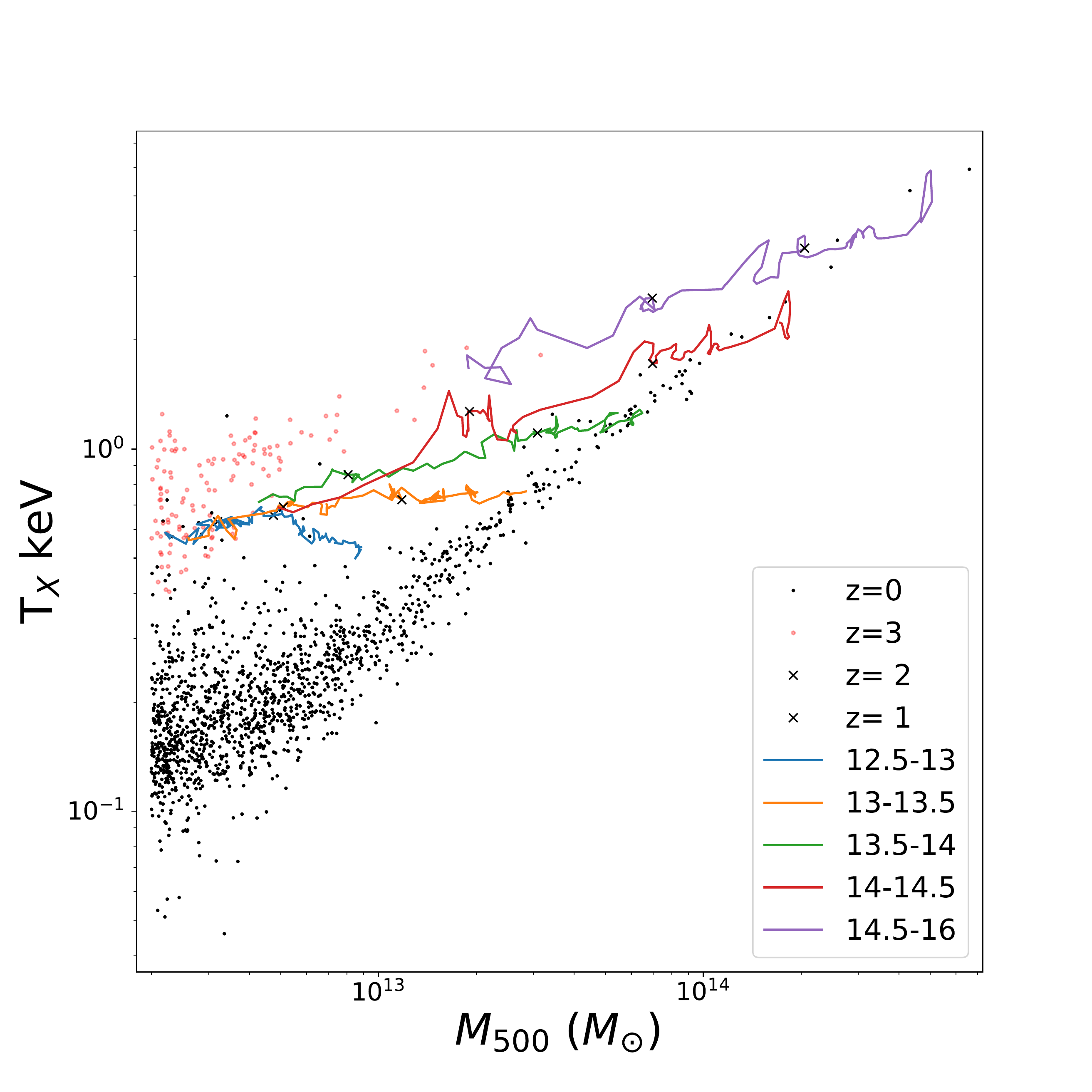}
    \caption{A plot showing the median evolution of a halo from $z= 3$ to $z= 0$, binned by final $M_{500}$ with blue, yellow, green, red, and finally purple representing least to most massive mass bins. X marks indicate the positions of $z=2$ and $z=1$ on each track. Light grey dots represent the scaling relation at $z=3$ while black dots represent the scaling relation at $z=0$. All but the least massive bin show a clear increase in $T_X$. All but the most massive bin seem to originate from the same area at $z=3$ suggesting there is no clear evolution by population based on $T_X$ alone.}
    \label{fig:ScalingTrack_M_T}
\end{figure}

Figure \ref{fig:M500_Tx_Scaling_Evolution} shows the $T_X-M_{500}$ scaling relation at $z=3,2.5,2,1.5,1,0.5,0$ with points colour-coded by $f_{BH}=M_{BH}/M_{500}$, similar to Fig.~\ref{fig:M500_Lx_Scaling_Evolution}.  The self-similar relation is again shown as the dashed line in each panel.

The evolution from $z=3\to 0$ shows a very slight steepening of the slope driven by a decrease of $T_X$ amongst the lowest mass halos. The most notable trend is that, amongst the lower mass halos, we see a large scatter in $T_X$ increase from a span of around 0.5~dex, to $>1$~dex from $z=3\to 0$, with again the most significant change happening around $z\sim 1.5$.

Looking at the trend with $f_{BH}$ reveals a muddier situation than with $L_X$.  At high redshifts, there is a clear (though mild) trend that high-$f_{BH}$ halos have high $T_X$.  This makes sense because the integrated AGN feedback energy is higher for high-mass black holes.  Yet, the trend disappears at $z\la 1$.  This suggest that the switching on of jet feedback is breaking the relation between $T_X$ and $M_{BH}$ amongst smaller halos.  A possible explanation is that jet feedback expels the hottest gas out of low-mass halos altogether, leaving only (relatively) cooler gas. As a result, there is not a simple relationship with $f_{BH}$ for the $T_X-M_{500}$ scaling; in \simba, stronger AGN feedback does not monotonically increase $T_X$.

Figure \ref{fig:ScalingTrack_M_T} shows median tracks within different halo mass bins, tracked backwards in time from the $z=0$ population, analogous to Fig.~\ref{fig:ScalingTrack_M_L}.  Again, we show 5 $z=0$ halo mass bins in $\log\ M_{500}$ as denoted in the legend, and we plot tick marks along the tracks at integer redshifts.  The tracks start at $z=3$ on the left hand side and evolve rightwards (growing in $M_{500}$) with time.

We see that the evolution here has some similarities but also some differences relative to the tracks in $L_X-M_{500}$ space.  The highest mass halos show a clear increase in temperature from $z=3\to 0$, while the least massive halos at $z=3$ show less change in the X-ray weighted gas temperature.  Interestingly, all halos start along the same mildly upwards trend, but while the massive halos continue that trend to $z=0$, 
at $z\la 1$ the low-mass halos stop evolving in $T_X$, with the lowest mass bin actually dropping in median $T_X$ since $z\sim 1$.  The net effect is a slight steepening of the $T_X-M_{500}$ slope.

This differential evolution with halo mass is likely driven by the interplay between jet feedback and structure formation.  At the lower masses, the jets are fast enough to drive substantial material out of halos, heating the surrounding gas but depositing less energy into the halo itself.  At the higher masses, the jets cannot escape the halo as easily, and the kinetic energy is thermalised within the halo.

\subsection{$f_{\rm gas}-M_{500}$ evolution}

\begin{figure}
     \centering
     \includegraphics[width = 0.5\textwidth, trim = 50 0 100 0, clip]{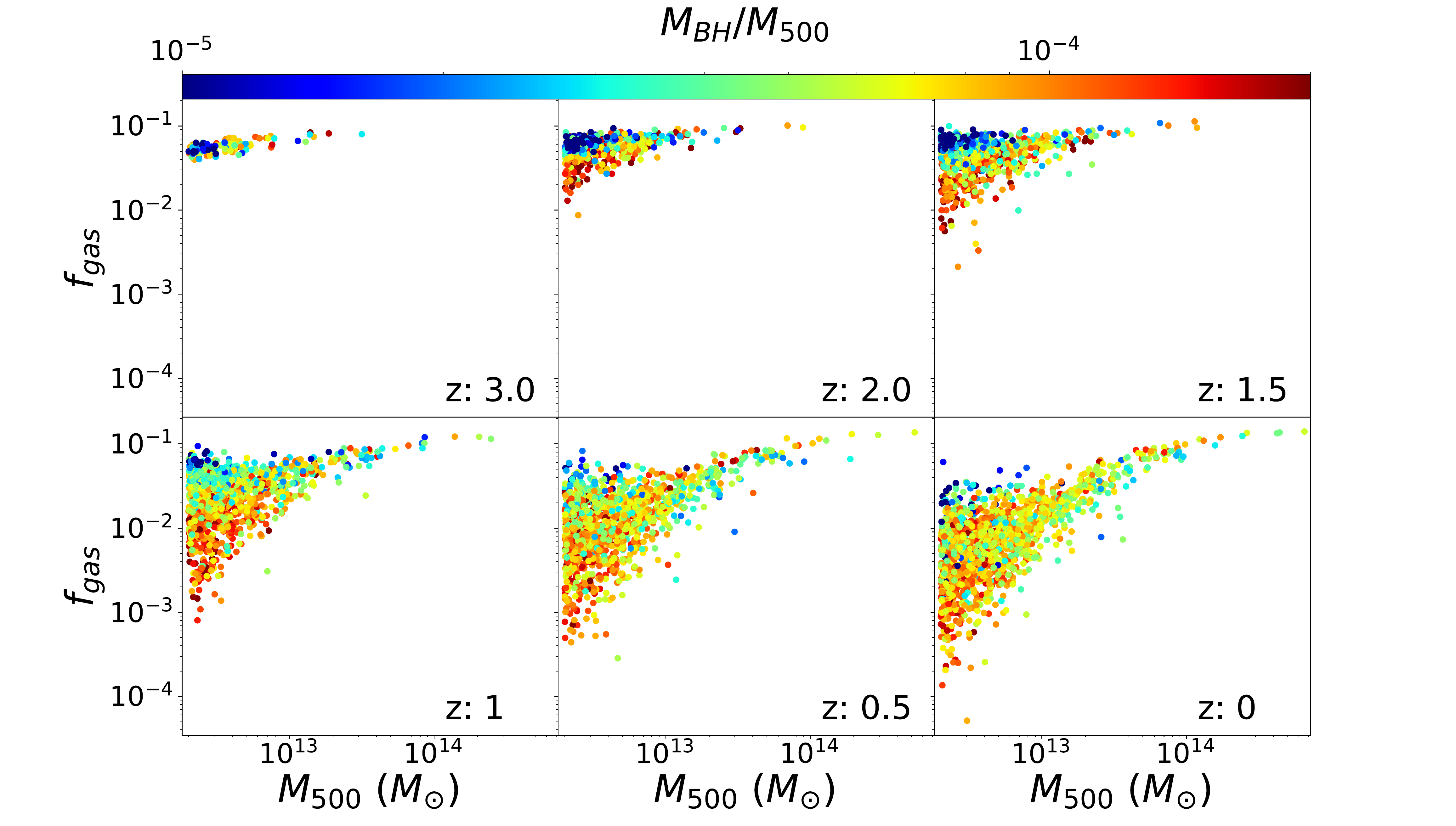}
     \caption{$f_{\rm gas}-M_{500}$ scaling relation from $z=3$ to $z=0$ at intervals of 0.5. The colourbar is scaled as $M_{BH}/M_{500}$ going from red to blue. A small evolution in scatter of the least massive halos is seen as $M_{500}$ increases up to $z=1.5$. After $z=1.5$ we see a large increase in scatter at the low mass end with a a significant decrease in hot gas mass fraction at these masses. This coincides with the switch on of jet feedback. Halos with $M_{500} < 10^{13.5} M_{\odot}$ show higher $M_{BH}$ corresponding to lower $f_{\rm gas}$.}
     
     \label{fig:M500_Gas_Scaling_Evolution}
\end{figure}

\begin{figure}
    \centering
    \includegraphics[width = 0.5\textwidth]{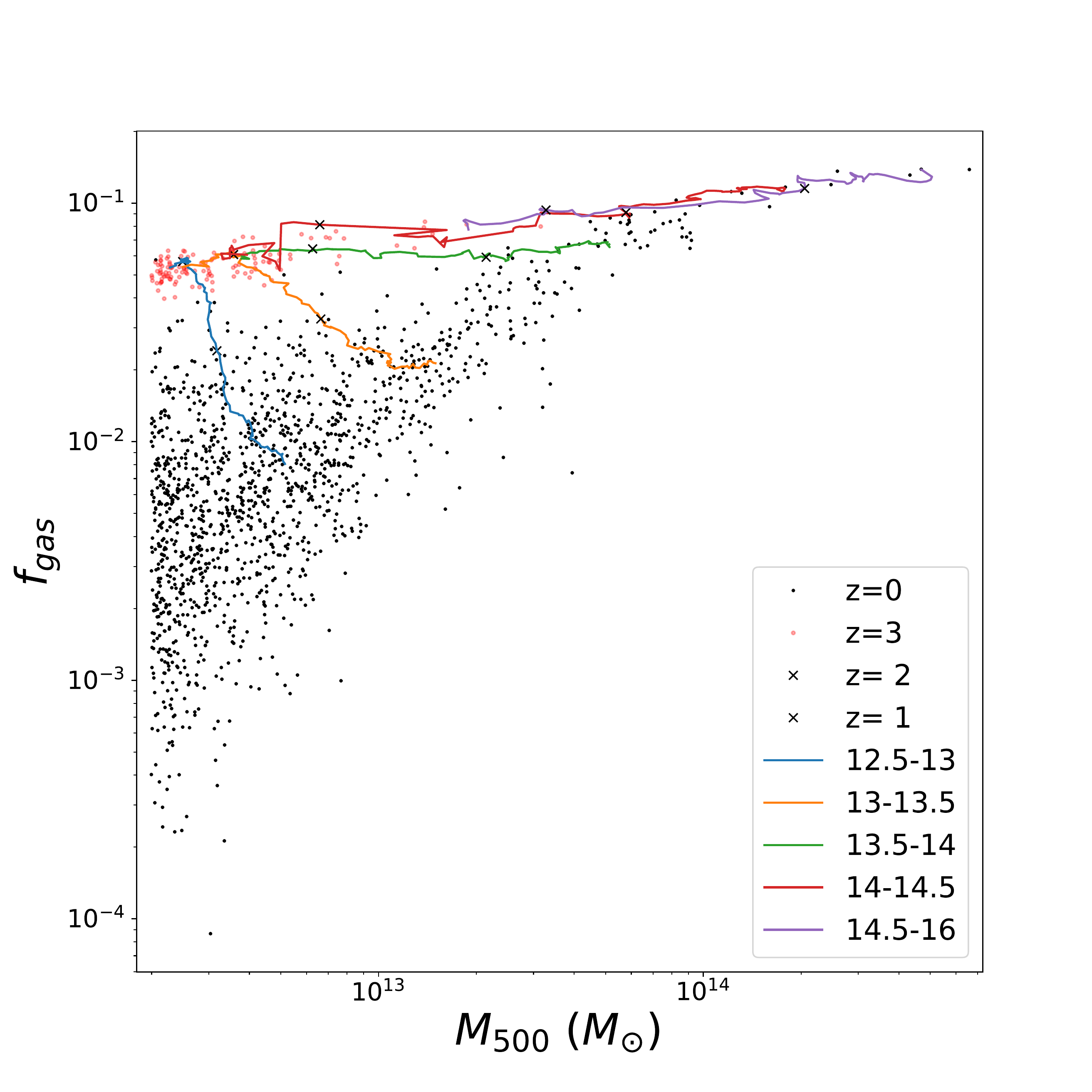}
    \caption{A plot showing the median evolution of a halo from $z=3$ to $z=0$, binned by final $M_{500}$ with blue, yellow, green, red, and finally purple representing least to most massive mass bins. X marks indicate the positions of $z=2$ and $z=1$ on each track. Light grey dots represent the scaling relation at $z=3$ while black dots represent the scaling relation at $z=0$. Halos with $M_{500} > 10^{13.5} M_{\odot}$ show a slight increase in $f_{\rm gas}$ as mass increases by around an order of magnitude. However, halos with $M_{500} < 10^{13.5} M_{\odot}$ show a clear drop in $f_{\rm gas}$ suggesting jet feedback is working to remove hot gas from the least massive halos.}
    \label{fig:ScalingTrack_M_G}
\end{figure}

In examining $L_X$ and $T_X$ vs. $M_{500}$, we have implicated the halo hot gas fraction $f_{\rm gas}$ as an important quantity that is directly impacted by jet feedback. \citet{Robson:2020} showed that without jets on, gas fractions show much larger values in small halos, with not nearly as strong a trend with halo mass.  Thus it is worth examining how the gas fraction evolves with time, to provide insights into the evolution of other more directly observable quantities.
 
Figure \ref{fig:M500_Gas_Scaling_Evolution} shows the $f_{\rm gas}-M_{500}$ scaling relation at $z=3,2.5,2,1.5,1,0.5,0$ with points scaled by the stellar mass fraction $f_{\rm star}$. We note that the top of each panel corresponds to $f_{\rm gas}$ reaching the cosmic halo baryon fraction $\Omega_b/(\Omega_m+\Omega_b)$.  We define hot gas as gas with $T>10^{5.5}$K. 

From $z=3\to 1.5$ we see that $f_{\rm gas}$ has a fairly weak trend with halo mass.  As time goes and particularly at $z\la 1.5$, the scatter increases dramatically at the low-mass end, with an increasing number of halos with lower $f_{\rm gas}$ values. This scatter introduced at lower masses results in a steepening of the $f_{\rm gas}-M_{500}$ scaling relation.  

The colour scaling shows an inverse relation between $M_{BH}$ and $f_{\rm gas}$ at $M_{500} < 10^{13.5} M_{\odot}$, with lower mass black holes corresponding to higher hot gas fractions.  This trend is already beginning at $z=2$, showing the effects of early jet feedback that results in early quenched galaxies, with the effects ramping up most strongly from $z=2\to 1.5$.  The lower hot gas content in halos with more AGN feedback might seem counterintuitive, since one expects AGN feedback to heat gas.  But the wind speed of the jets is such that it can easily unbind gas at these lowest masses, whereas for the high-mass halos it will be more easily trapped within the halo.  Thus the impact of AGN jets isn't so much to heat the halo gas, as expel it.  This is why we see a correspondingly strong trend with $f_{BH}$ in $L_X$ which is sensitive to gas density, but a weak trend with $T_X$. We will examine gas densities later when we look at the X-ray profiles.

Figure \ref{fig:ScalingTrack_M_G} tracks the gas fraction over time for these halos.  We see that it is only the less massive halos showing a decrease in $f_{\rm gas}$ with halos of $M_{500} > 10^{13.5} M_{\odot}$ demonstrating steady or mildly increasing $f_{\rm gas}$.  This trend is qualitatively similar to what was seen in tracking $L_X$ (Fig.~\ref{fig:ScalingTrack_M_L}), highlighting the direct impact of $f_{\rm gas}$ on $L_X$.

\subsection{$Z_X-M_{500}$ evolution}

\begin{figure}
     \centering
     \includegraphics[width = 0.5\textwidth, trim = 50 0 100 0, clip]{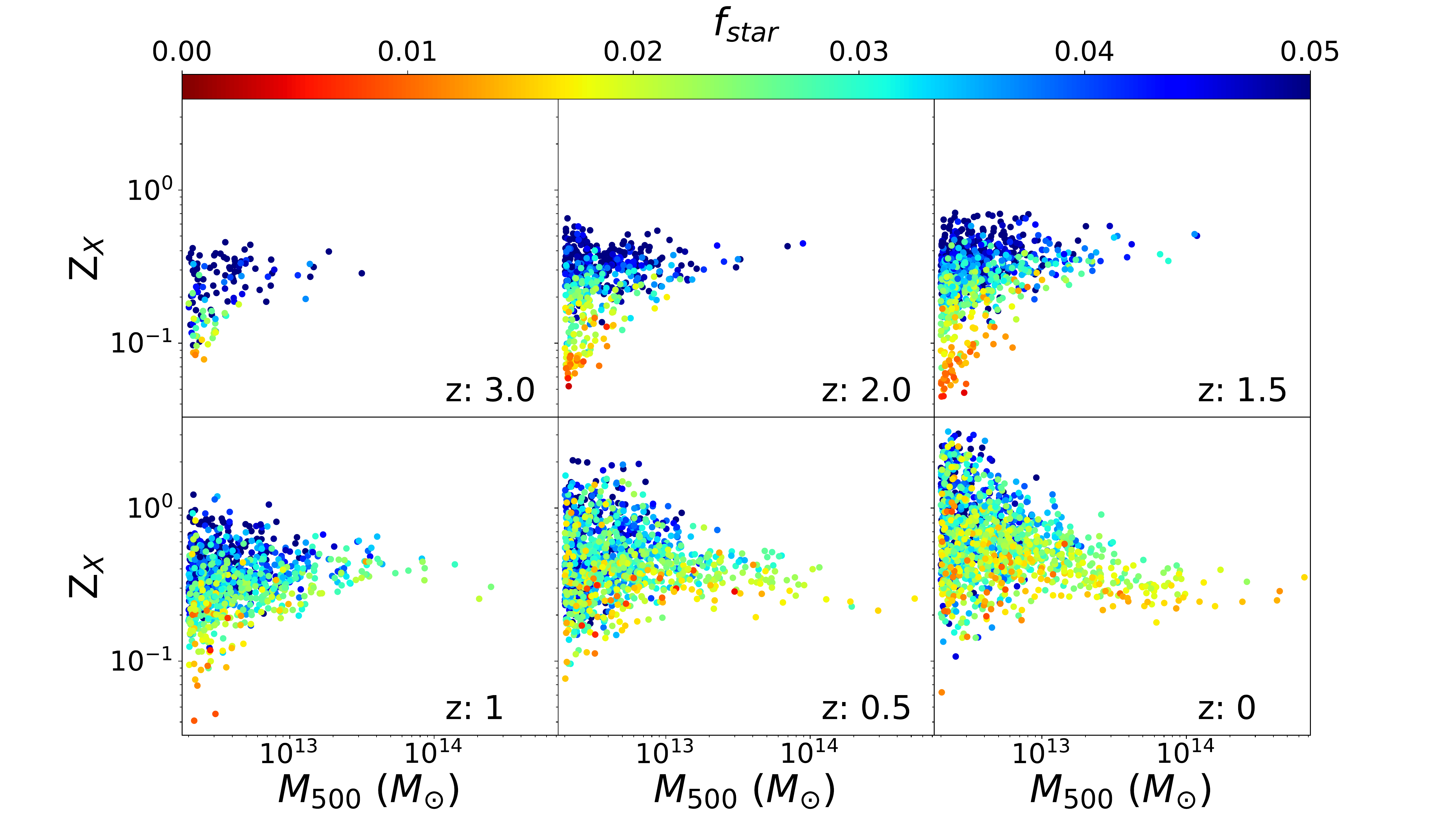}
     \caption{$Z_X-M_{500}$ scaling relation from $z=3$ to $z=0$ at intervals of 0.5. The colourbar is scaled as $f_{star}$ from 0 to 0.1 going from red to blue. A clear increase in metallicity is seen at the low mass end.  }
     
     \label{fig:M500_Zx_Scaling_Evolution}
 \end{figure}
 
While emission via bremsstrahlung dominates for systems with $T_X \ga 3 keV$ metal line emission becomes increasingly important towards lower-$T_X$ systems. This includes all but the most massive systems in our sample, and as such an examination of the hot gas metallicity of these systems also provides insight.  Moreover, metals provide a unique tracer of supernova-generated products and their distribution into intragroup gas.

Figure                        \ref{fig:M500_Zx_Scaling_Evolution} shows the $Z_X-M_{500}$ scaling relation at        $z=3,2.5,2.0,1.5,1.0,0.5,0$ with points scaled by the stellar mass fraction $f_{star}$. $f_{star}$ here is defined as the ratio of stellar mass within $R_{500}$ against $M_{500}$. Through $f_{star}$ we can infer a level of stellar feedback within a cluster, a key driver behind metallicity. Scaling by $M_{BH}/M_{500}$, as in previous plots, showed no relation motivating us to investigate the effects of different forms of feedback on evolution. 

Between $z=3$ and $z=1.5$ little evolution is seen in the shape of the relation, with an increase in halos above the mass cut we see little change in metallicity at any given mass with the exception of more massive halos become present. However after $z<1.5$ we begin to see a change in metallicities seen among low mass halos. Much like previous scaling relations, we see an increase is seen in the scatter at lower masses. In particular, the lowest mass groups can show high intragroup metallicities, exceeding solar in some cases at $M_{500}<10^{13}M_\odot$. 

Noting the colour-coding, the high intragroup metallicities correlate fairly well with high $f_{star}$ in the central galaxy. This trend is very pronounced at high-$z$, but remains noticeable all the way to $z=0$. At $z>1$, the lowest metallicity systems have very low stellar fractions, which suggests that these systems have been quenched early on~\citep{Cui:2021}.   This is perhaps not surprising, since the stars are responsible for producing much of the metals in the system, particularly early on.  Moving to lower redshifts, re-accretion of enriched intergalactic gas~\citep[i.e. ``outside-in enrichment";][]{Oppenheimer:2012} imparts a metal floor in these high density environments, and dilutes the strong trend arising purely from self-enrichment.  Nonetheless \simba\ predicts that poor groups with higher stellar masses~\citep[which also tend to have more star-forming galaxies;][]{Cui:2021} will generally to have higher metallicities.

\subsection{$L_X-T_X$ evolution}

\begin{figure}
     \centering
     \includegraphics[width = 0.5\textwidth, trim = 50 0 100 0, clip]{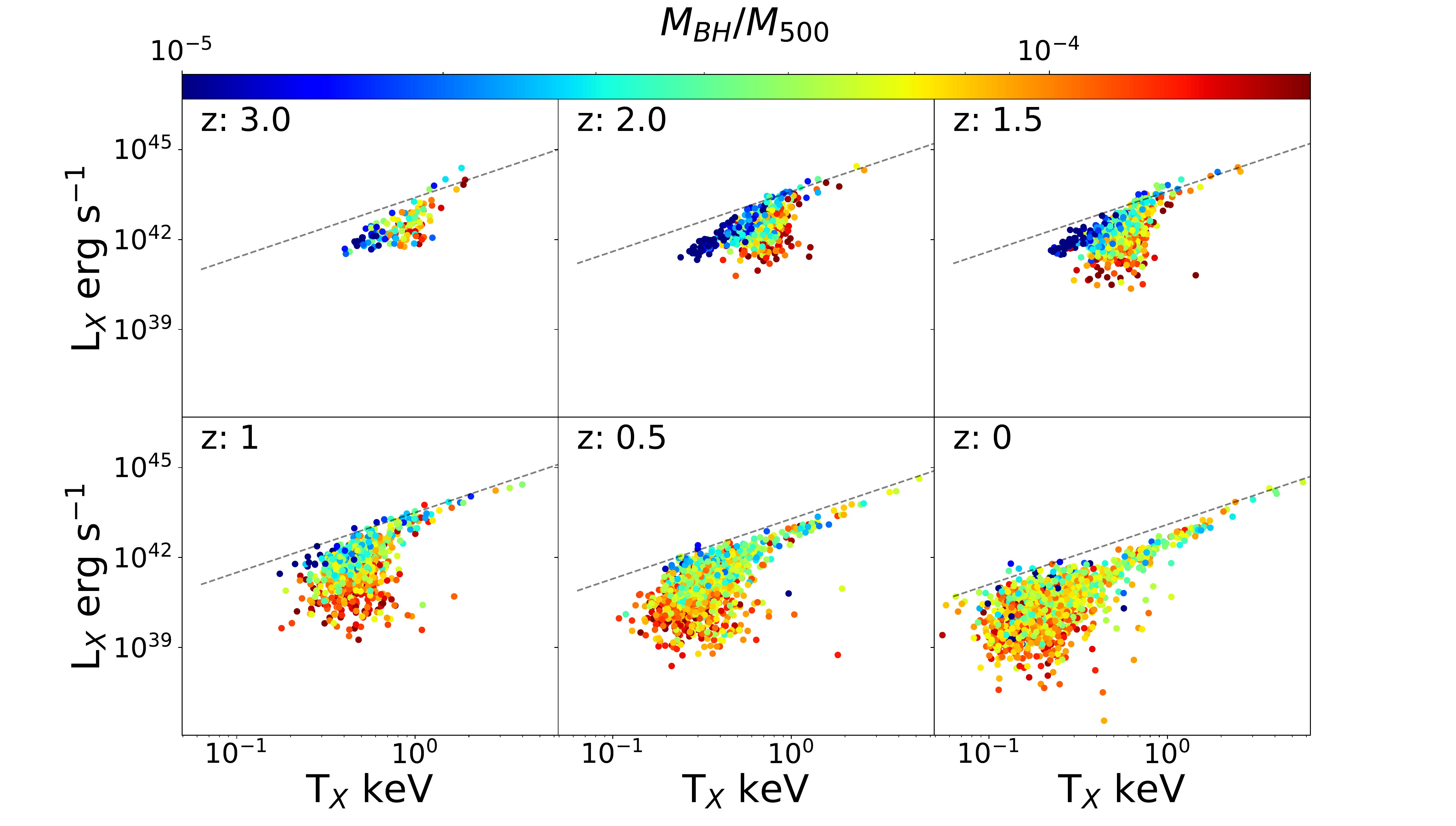}
     \caption{$T_X-L_X$ scaling relation from $z=3$ to $z=0$ at intervals of 0.5. The colourbar is scaled as $M_{BH}/M_{500}$. The expected self-similar scaling normalise to the most massive halos is shown by the dashed line. A tight scaling is retained amongst the hottest most luminous halos, however a large scatter is introduced at low temperatures and luminosities. The introduction of jet feedback after $z=1.5$ coincides with the introduction of a number of low temperature, low luminosity halos exhibiting a large scatter. }
     
     \label{fig:Lx_Tx_Scaling_Evolution}
\end{figure}

The $L_X-T_X$ relation has long been a key observational diagnostic of hot gas in massive halos. The evolution of this relation will obviously be a combination of the evolution in each quantity versus $M_{500}$ from the previous sections. Here we examine this evolution directly, including examining how groups at different masses evolve within $L_X-T_X$ space.

Figure \ref{fig:Lx_Tx_Scaling_Evolution} shows the $L_X-T_X$ scaling relation at $z=3,2.5,2,1.5,1,0.5,0$ with points colour-coded by $M_{BH}/M_{500}$.  We impose a mass cut of $M_{500}>10^{12.3}M_\odot$ at each redshift.  The dashed line shows the self-similar relation normalized to the highest $T_X$ group.

From $z=3\to 1.5$, the $L_X-T_X$ relation looks broadly very similar. The largest halos with the highest $T_X$ are already on the self-similar relation, and by $z\sim 1$ there is a fairly tight locus which sets an upper edge following self-similarity.  Over time, there is a growing scatter of points to lower $L_X$.  This becomes particularly prominent at $z\sim 1-1.5$ in the lower mass halos.  We have similarly seen in $L_X-M_{500}$ and $T_X-M_{500}$ scaling relations that there is a change in evolution for lower-mass groups around this epoch.

There is a marked trend that the groups with the lowest $L_X$ values have the highest $f_{BH}$ values.  This directly implicates the impact of AGN feedback in growing a population of low-mass groups that deviates from self-similarity.  As we discussed earlier, the effect of the jets is to evacuate the low-mass halos. This lowers the amount of X-ray emitting gas and thus $L_X$, but we showed that it does not have a strong impact on $T_X$ values in low-mass systems because there is the competing effect that the hottest gas is evacuated from these systems altogether.  \simba's prediction of a strong anti-correlation between $L_X$ and $f_{BH}$ at $T_X\la 1$~keV can be tested with next-generation X-ray facilities.

\begin{figure}
    \centering
    \includegraphics[width = 0.5\textwidth]{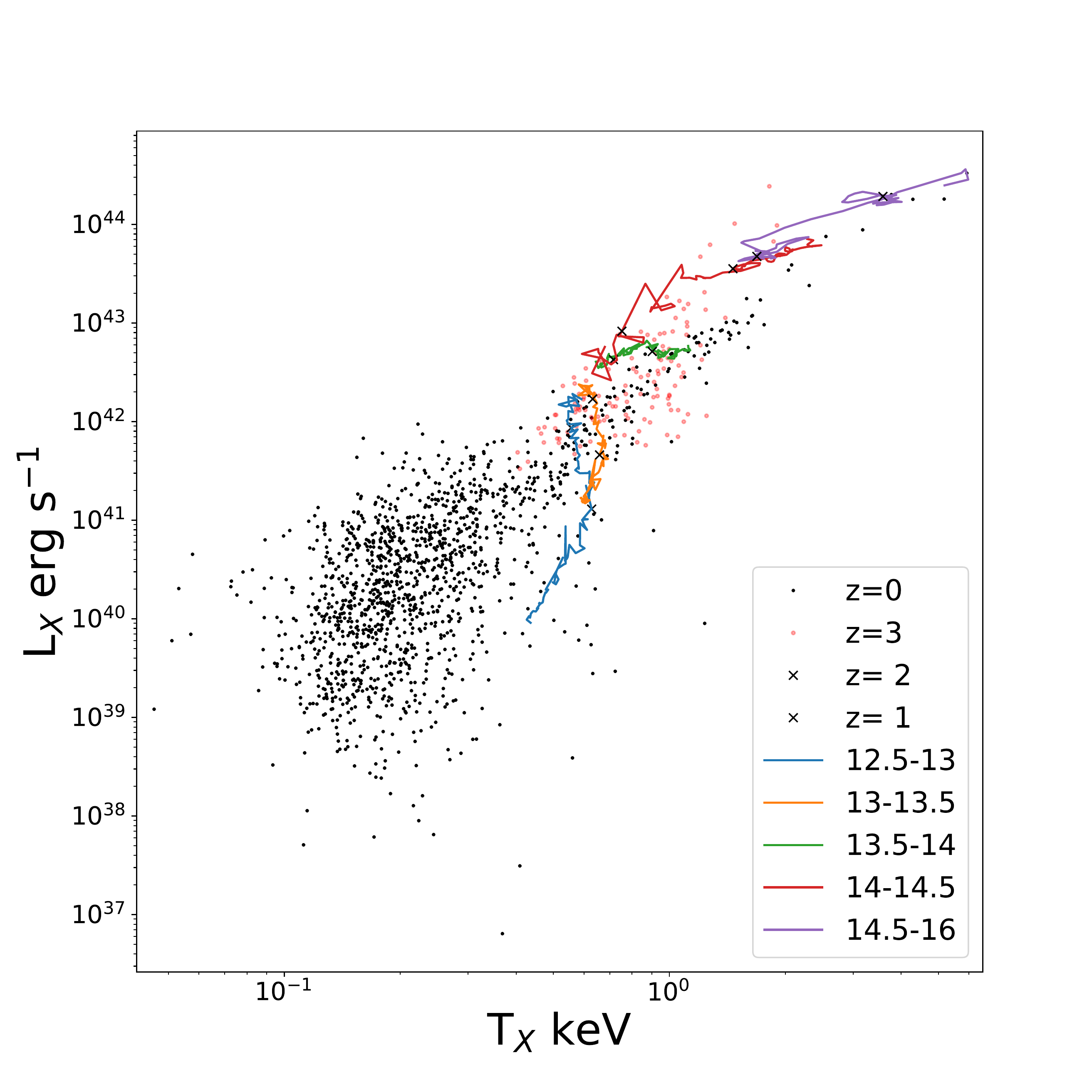}
    \caption{A plot showing the median evolution of a halo from $z=3$ to $z=0$, binned by final $M_{500}$ with blue, yellow, green, red, and finally purple representing least to most massive mass bins. X marks indicate the positions of $z=2$ and $z=1$ on each track. Light grey dots represent the scaling relation at $z=3$ while black dots represent the scaling relation at $z=0$. Halos with $M_{500} > 10^{13} M_{\odot}$ show little evolution in the $L_X-T_X$ space however the least massive halos below this threshold demonstrate a drop of nearly 2 orders of magnitude in $L_X$ while maintaining a consistent $T_X$.}
    \label{fig:ScalingTrack_L_T}
\end{figure}

In Figure~\ref{fig:ScalingTrack_L_T} we show the evolution of the $L_X-T_X$ relation in 5 $z=0$  $\log_{10}M_{500}$ mass bins as indicated in the legend, as before.  In this case, the red points depicting $z=3$ values are not wholly distinct from the $z=0$ black points, since there is not a monotonic increase in $T_X$ as there is in $M_{500}$.

The scaling tracks offer a very interesting perspective on $L_X-T_X$ evolution.  The overall relation at $z=3$ and $z=0$ are not that different in terms of the slope and amplitude, although at $z=0$ there are many more low-$T_X$ systems with a large scatter in $L_X$.  The apparent lack of evolution in the $L_X-T_X$ relation has been noted in some observations of clusters \citep{Holden:2002,Borgani:2001}, though not for lower-mass groups since current X-ray data lacks the sensitivity at $z\gg 0$.  Several other studies have shown this weak evolution at $z>0.3$ however show varying levels of evolution from $z=0.3 \to z=0$ motivated by the consideration of the effect of cool core clusters \citep{Vikhlinin:2002, Ettori:2004}. 

The modest evolution in $L_X-T_X$ is seen across all masses, though again there is a different trend for $M_{500}\ga 10^{14}M_\odot$ systems that tend to evolve  rightwards with a modest increase in $L_X$, while $M_{500}\la 10^{13.5}M_\odot$ evolve downwards in $L_X$ with little change in $T_X$.  The net effect is to leave the overall relation relatively unchanged, but to spread out the groups along the relation.  Hence even though $L_X-M_{500}$ and $T_X-M{500}$ individually show significant change over time, the $L_X-T_X$ relation combines these two coincidentally into a nearly non-evolving relation, albeit with a large increase in scatter at very low $T_X$.

\section{Effects of Feedback on Scaling Relations}

We have implicated jet feedback in driving the evolution of the intragroup medium scaling relations particularly at the low-$M_{500}$ end.  We can test this more directly by examining a suite of smaller $50\hmpc$ runs having the same resolution, with individual AGN and star-formation modules turned on and off.   This offers us the chance to directly connect the scaling relations to the physics of the implemented feedback processes.

As previously described, the variant models used for comparison are the full \simba, No-X, No-Jet, No-AGN, and No-Feedback models. No-X turns off only X-ray feedback, No-Jet additionally turns off jet feedback, No-AGN additionally turns off radiative AGN feedback, and No-Feedback additionally turns off star formation winds. Overall, the No-AGN results were very similar to the No-Jet results, showing that the radiative AGN feedback has minimal effect on X-ray scaling relations~\citep[as likewise found for galaxy properties;][]{Simba}, and as such it has been left out of the following plots for brevity. By investigating the evolution of scaling relations with each feedback variant run we are able to demonstrate each specific model's impact on each of the previously studied scaling relations.

\subsection{Hot Gas Fraction}

\begin{figure}
    \centering
    \includegraphics[width = 0.5\textwidth]{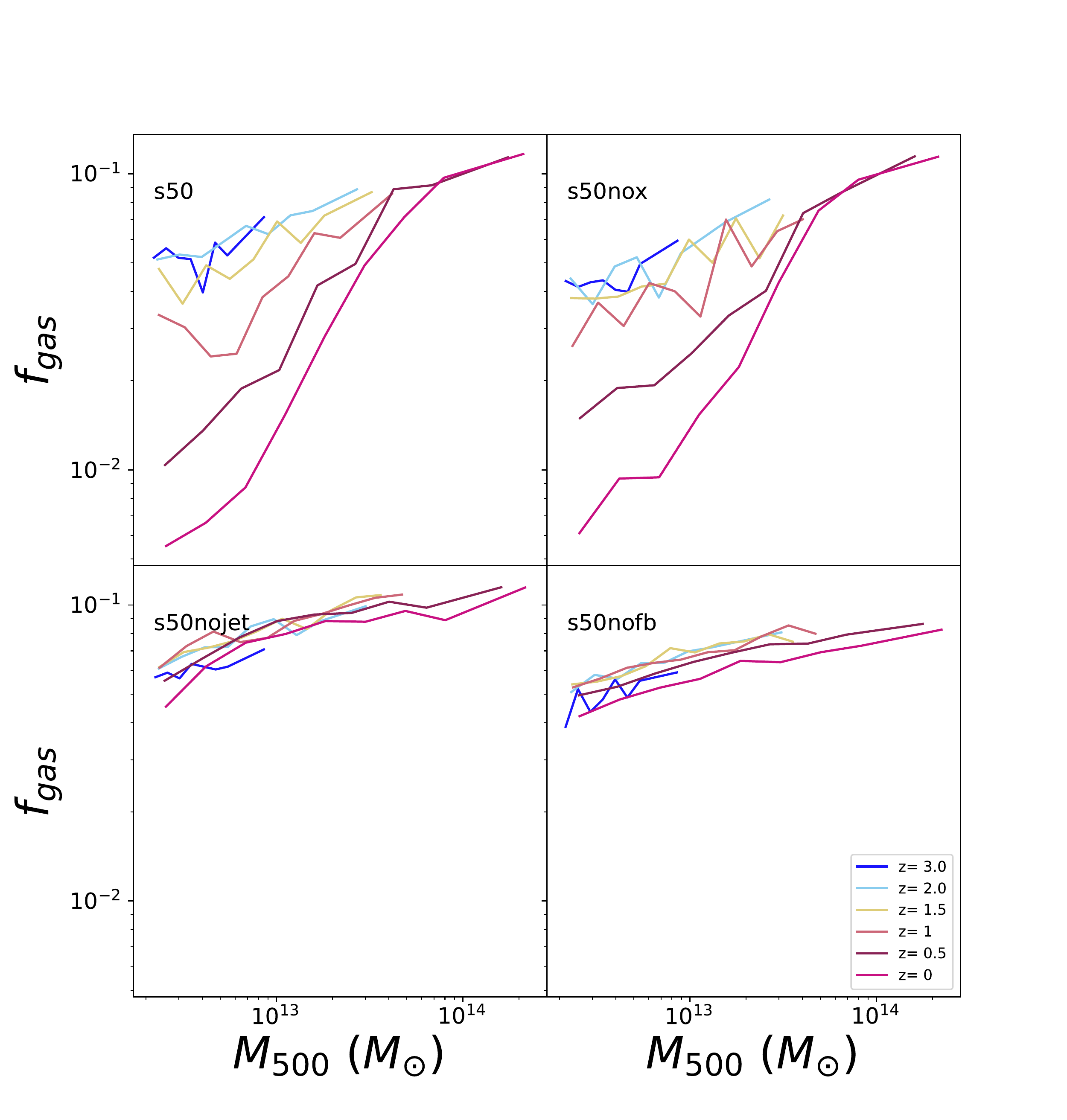} 
    \caption{Evolution from $z=3$ to $z=0$ of the running median of the $f_{\rm gas}-M_{500}$ scaling relation for \simba, \simba without X-ray feedback, \simba without jet feedback, and \simba with no feedback. All models lacking jet feedback (No-Jet, No-AGN, No-Feedback) show no evolution in $f_{\rm gas}$ while both models including jet feedback (\simba, No-X) exhibit a very large drop in $f_{\rm gas}$ in the halos with $M_{500} < 10^{13.5} M_{\odot}$.}
    \label{fig:M_fgas_comp}
\end{figure}

Since the hot gas fraction is implicated in driving much of the evolution of the scaling relations, we begin by examining how feedback impacts this quantity.

Figure \ref{fig:M_fgas_comp} shows the median $f_{\rm gas}$ as a function of $M_{500}$ at redshifts $z=3\to 0$ (blue to red) in various colours as indicated in the legend.  The upper left panel shows a $50\hmpc$ run with full \simba\ physics, and the other panels show the No-X run (upper right), No-Jet (lower left), and No-Feedback (lower right).

There is a dramatic difference in the evolution of $f_{\rm gas}$ once the jet feedback is turned on, seen by comparing No-Jet in the lower left, and No-X in the upper right.  Turning on star formation feedback and X-ray AGN feedback have in contrast a fairly minor effect.  We clearly see the impact of jet feedback at $M_{500} < 10^{13.5} M_{\odot}$ at $z\la 1.5$ while having a limited effect at $M_{500} \ga 10^{13.5} M_{\odot}$. Once jets are turned off (No-Jet and No-Feedback runs), there is essentially no evolution in $f_{\rm gas}$.  This clearly demonstrates that it is jets that are responsible for evacuating the hot gas from halos.  

The most notable evolution occurs at $z\sim 1-1.5$, which is consistent with the strong increase in scatter seen in various scaling relations at these lower masses around this epoch.  This suggests that jet feedback drives gas out of halos in a more stochastic manner, thereby creating large scatter in $L_X$.  Since jet feedback becomes strong at $\fedd$ below a few percent, which tends to occur when the black hole mass grows to above $\ga 10^8M_\odot$ \citet{Thomas:2019}, this then connects the larger black holes with the lower $f_{\rm gas}$ and hence lower $L_X$.  We can see this more clearly by examining the $L_X$ and $T_X$ relations more directly using our feedback variant runs, which we do next.

\subsection{$L_X - M_{500}$}

\begin{figure}
    \centering
    \includegraphics[width = 0.5\textwidth]{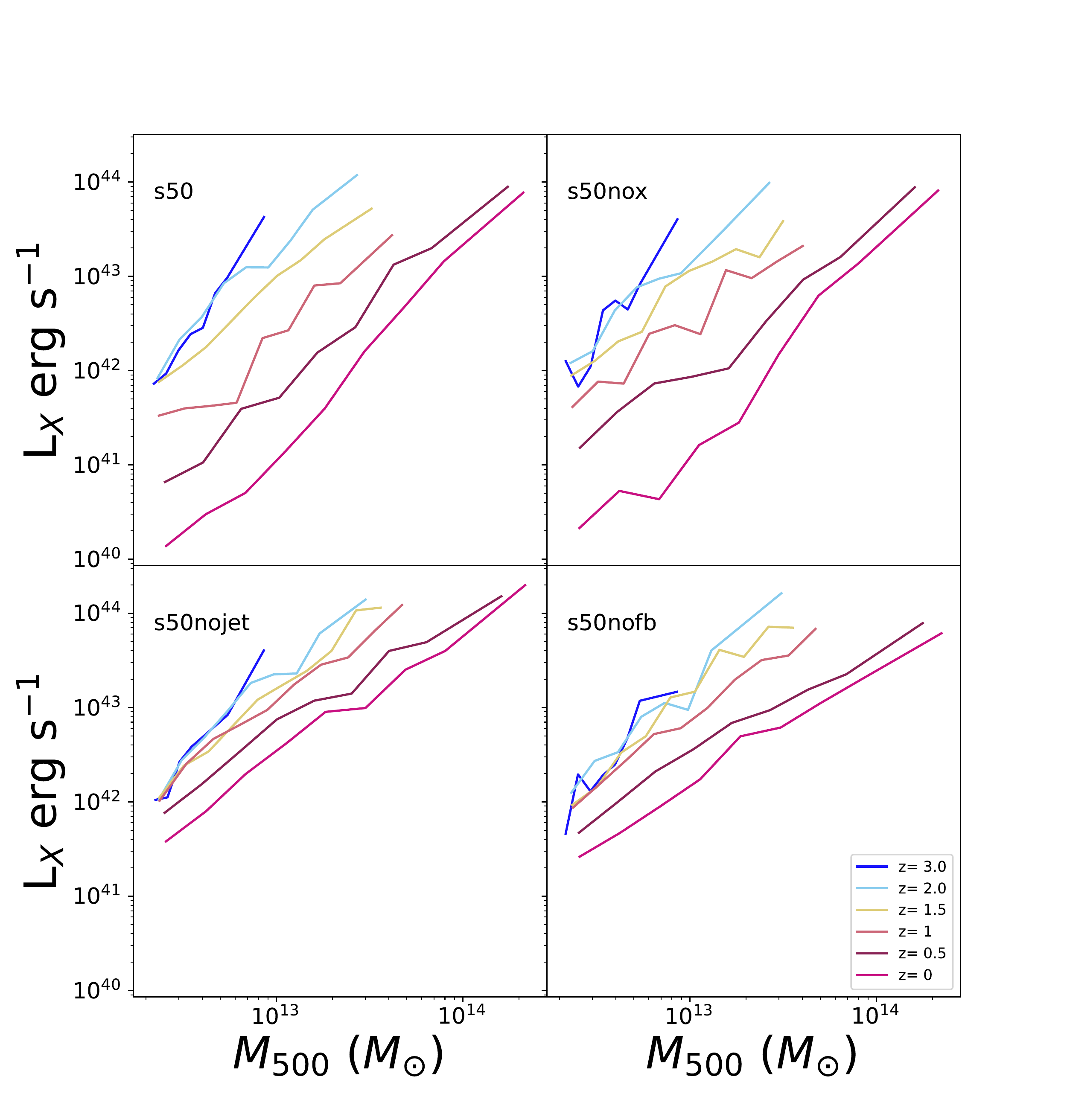}
    \caption{Evolution from $z=3$ to $z=0$ of the running median of the $L_X-M_{500}$ scaling relation for \simba, \simba without X-ray feedback, \simba without jet feedback, and \simba with no feedback. Little change is seen between the No-Feedback, No-AGN, and No-Jet runs, with the No-X and \simba runs exhibiting similar larger evolutions. The key difference here being the inclusion of jet feedback driving the evolution of the scaling relation.}
    \label{fig:ML_Z_comp}
\end{figure}
Figure \ref{fig:ML_Z_comp} shows the median $L_X$ as a function of $M_{500}$ at reshifts $z = 3 \to 0$ (blue to red) in various colours as indicated in the legend. The upper left panel shows a $50\hmpc$ run with full \simba physics, and the other panels show the No-X run (upper right), No-Jet (lower left), and No-Feedback (lower right), as in Figure~\ref{fig:M_fgas_comp}.

In full-physics \simba we see a steady decrease in the lowest luminosities, with the highest luminosities remaining relatively consistent as masses increase from $10^{13} M_{\odot}$ to $>10^{14} M_{\odot}$. This is inline with what we saw in the $100\hmpc$ box. We see very little difference in the overall evolution of the No-X run, suggesting that X-ray feedback has minimal effect on the evolution of the X-ray luminosity. 

In contrast, in the No-Jet run we immediately see the impact of jet feedback. Scaling relations at high redshifts match closely with \simba's, but much less evolution in $L_X$ is seen, with minimum values not going below $10^{41}$~erg~s$^{-1}$.  The slope of the relation is mostly invariant.  NoFB shows a similarly low amount of evolution, showing that SF feedback (and radiative AGN feedback) have fairly minimal impact on this scaling relation.  However, there are some non-trivial differences, particularly in the most massive halos that seem to have lower $L_X$ without any feedback.  Hence stellar and radiative AGN feedback seem to work to increase the X-ray luminosities.  The origin is likely from the slight increase in hot gas content from NoFB$\to$No-Jet as seen in Figure~\ref{fig:M_fgas_comp}, because the NoFB case locks more of the halo's baryons into stars~\citep{Appleby:2021}.
 
\begin{figure}
    \centering
    \includegraphics[width=0.5\textwidth]{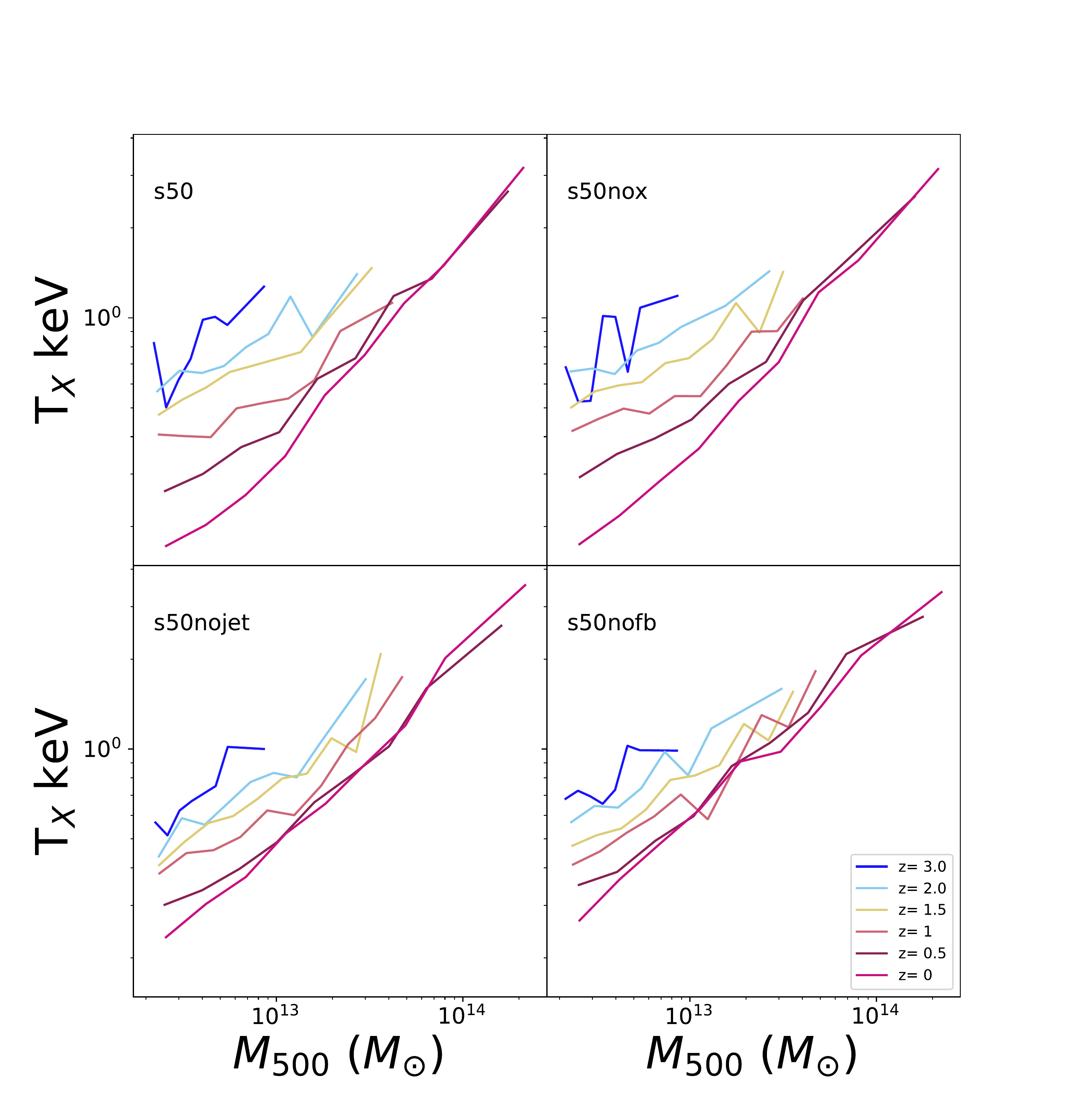}
    \caption{Evolution from $z=3$ to $z=0$ of the running median of the $T_X-M_{500}$ scaling relation for \simba, \simba without X-ray feedback, \simba without jet feedback, and \simba with no feedback. Both \simba and No-X show similar evolutions while No-Jet, No-AGN, and No-Feedback show much tighter evolutions indicating jet feedbacks effect in lower $T_X$ in low mass halos. No noticeable difference is seen in the scaling relations of the most massive halos.}
    \label{fig:MT_Z_comp}
\end{figure}
\subsection{$T_X - M_{500}$}
Figure \ref{fig:MT_Z_comp} shows the median $T_X$ as a function of $M_{500}$ at redshifts $z = 3 \to 0$ (blue to red) in various colours as indicated in the legend. The panels are ordered among the feedback variants as before. 

We see quantitatively similar behaviour as in the $L_X-M_{500}$ relation, though less dramatically.  Jet feedback has the clearest impact on this scaling relation, particularly on the least massive halos.  This can be seen from the difference between the No-Jet in the lower left, and No-X in the upper right.  The noticeable difference in the evolution between these two runs begins at $z\sim 1.5$.  The outcome of jets is to, perhaps counter-intuitively, lower the temperature of the hot gas in low-mass halos; we speculated this earlier, but here it is clearly demonstrated.  At high masses, none of the feedback models have a significant effect on $T_X$.  Nonetheless, it is worth noting that stellar feedback impacts the temperature of low mass at early times, producing a steeper scaling relation. Thus $T_X-M_{500}$ evolution owes to a combination of different feedback processes operating at different times.

\subsection{$L_X-T_X$}

\begin{figure}
    \centering
    \includegraphics[width = 0.5\textwidth]{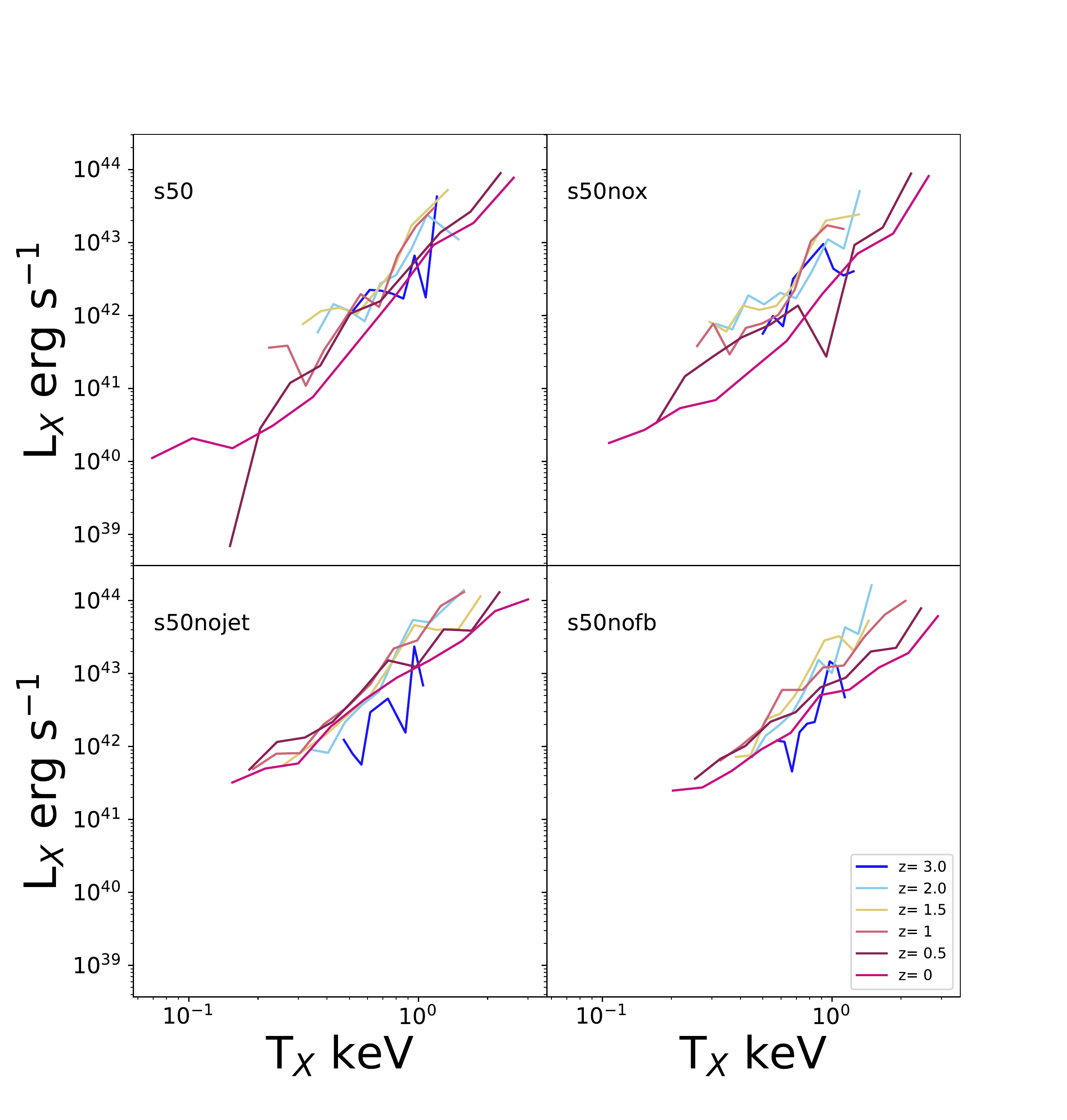}
    \caption{Evolution from $z=3$ to $z=0$ of the running median of the $T_X-L_X $ scaling relation for \simba, \simba without X-ray feedback, \simba without jet feedback, and \simba with no feedback. Little difference is seen in the NoFB, No-AGN, and No-Jet runs. The No-X run shows an further evolution into the low $T_X-L_X$ region of the plot indicating jets impact on both lower temperatures and luminosities. The \simba run shows even further evolution into the lower temperatures indicating that X-ray feedback may have a small effect in these low temperature halos at late times.}
    \label{fig:Tx_Lx_comp}
\end{figure}

Figure \ref{fig:Tx_Lx_comp} shows the median $T_X$ as a function of $M_{500}$ at redshifts $z = 3 \to 0$ (blue to red) in our various feedback runs.

Remarkably, despite clear evolution in $L_X-M_{500}$ and $T_X-M_{500}$, the $L_X-T_X$ relation seems to show little evolution, and indeed not much sensitivity to feedback.  This occurs because jet feedback lowers both $L_X$ and $T_X$, resulting in halos moving further down the $L_X-T_X$ relation, but not changing its slope or amplitude much.  The net effect is that AGN jet feedback ``stretches out" the $L_X-T_X$ relation, but doesn't substantially alter it.  In detail, the $L_X-T_X$ relation does get a bit steeper once jets are turned on, because the drop in $L_X$ due to the evacuation of halos is more severe than the lowered $T_X$ owing to the hot gas being unbound from low-mass halos.  Hence AGN jets do break the self-similarity in $L_X-T_X$ at low masses, but the effect is not as dramatic as seen in the individual relations vs. $M_{500}$.

In \citet{Robson:2020} we proposed that \simba's scaling relations at $z=0$ were a result of the implemented jet feedback evacuating hot gas in lower mass halos. In Fig \ref{fig:ML_Z_comp}  and Fig \ref{fig:MT_Z_comp}, \simba and No-X show a clear deviation in their evolution after $z=1.5$, around the time jet feedback kicks in in the \simba simulations. This lends credence to the idea that jets are responsible for \simba's ability to match X-ray scaling relations, which is mainly driven by the gas fraction evolution.

\subsection{$Z_X-M_{500}$}

\begin{figure}
    \centering
    \includegraphics[width =0.5\textwidth]{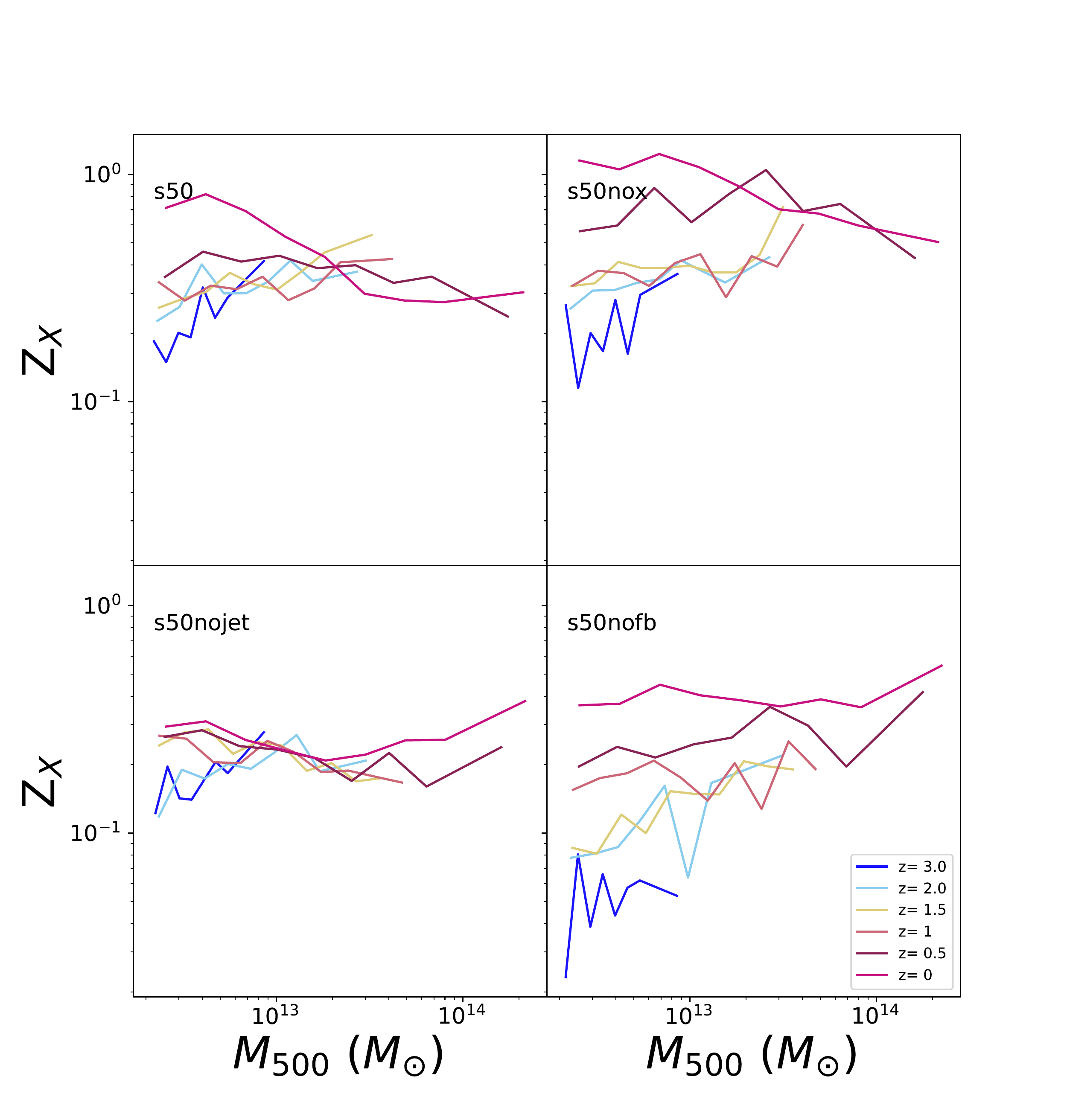}
    \caption{Evolution from $z=3$ to $z=0$ of the running median of the $Z_X-M_{500}$ scaling relation for \simba, \simba without X-ray feedback, \simba without jet feedback, and \simba with no feedback. The no feedback run shows a steady increase in $Z_X$ with redshift. With the introduction of stellar feedback in the no jet run we see no evolution from $z=3$ to $z=0$ with stellar winds raising the metallicity to $\approx 0.3$ before $z=3$. The no xray run shows jets increasing $Z_X$ at late times, especially amongst small halos. Finally the full \simba again shows very little evolution, demonstrating X-ray feedback's effect in full quenching halos in \simba.}
    \label{fig:Zx_M500_comp}
\end{figure}

Figure~\ref{fig:Zx_M500_comp} compares the $Z_X-M_{500}$ scaling relations of the various feedback runs, in the same format as the previous plots.

In contrast to the other quantities, jet feedback appears not to be the main driver of evolution in $Z_X$. In NoFB we see a steady increase in metallicity over time, which disappears in the No-Jet case, which instead has metallicities remaining fairly steady $Z_X\approx 0.2-0.3$ solar across time (i.e. roughly the $z=0$ value in the No-Feedback run).  We do not show it, but the No-AGN case is similar to No-Jet.  This shows us that stellar feedback is increasing the metallicity of hot gas to observed levels early on and then maintaining them. 

Nonetheless the jets do have some effect.  In the No-X run, with the introduction of jet feedback we see the X-ray weighted metallicities increase mildly after $z=1.5$ in the most massive halos relative to No-Jet, and slightly later but more noticeably after $z=1$ in the least massive halos.  Hence jets are an important factor for increasing the metallicity in hot intragroup gas at later epochs, presumably by transporting metals out of galaxies in the process of quenching them.

The inclusion of X-ray feedback in the \simba run then sees these late $Z_X$ brought back down to $\approx 0.3$ in all but the least massive halos at $z=0$.  This drop relative to No-X is concordant with the drop in the stellar fraction produced in massive galaxies~\citep{Simba}, thus indicates the effect of X-ray feedback in fully quenching galaxies in \simba.

Overall, jet feedback has the largest impact in altering the scaling relations from the No-Feedback case.  The No-Feedback scalings are, not surprisingly, broadly consistent with self-similar scaling, albeit with deviations at low masses owing to excessive cooling in this model~\citep{Dave:2002}.  Jets evacuate low mass halos and result in lower temperatures for the gas that remains, giving lower $L_X$ and $T_X$ values.  The metallicity, meanwhile, is primarily deposited by star formation feedback at early epochs, though jet feedback tends to result in higher $Z_X$ in low-mass halos at late epochs.

\section{Profiles}

\begin{figure*}
    \centering
    \includegraphics[width = \textwidth]{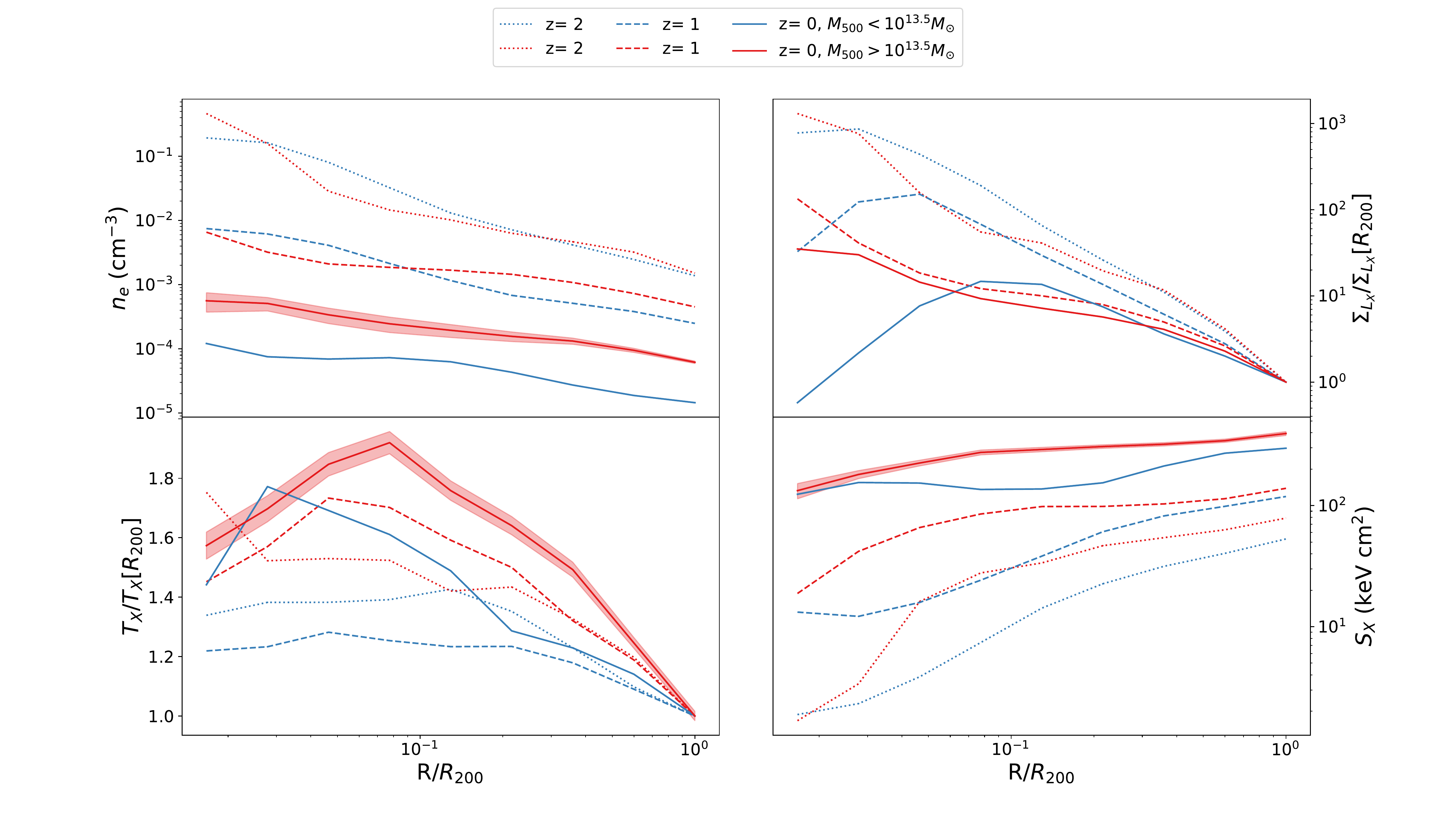}
    \caption{Median Profiles at $z= 2,1,0$ for electron density, $\Sigma_{L_X}$, $T_X$, and $S_X$. $z=2$ is represented by a dotted line, $z=1$ by a dashed line, and $z=0$ by a solid line. The median profiles of halos with $M_{500} < 10^{13.5} M_{\odot}$ are coloured blue while those $> 10^{13.5} M_{\odot}$ are represented by the red lines. The $\Sigma_{L_X}$ profiles are scaled to their values at $R_{200}$ to accentuate the evolution in shape however the shift in trends can still be seen in the other plots. $n_e$ shows a profile decreasing with radius at all redshifts, with profiles flattening due to a higher decrease in the core regions as we move from $z=2$ to $z=0$. We also see overall higher densities at all radii within the $M_{500} > 10^{13.5}$ bin at $z=0$, a shift from the similar levels at previous redshift. $\Sigma_{L_X}$ profiles for $M_{500} > 10^{13.5}$ show a slight flattening with redshift. This evolution matches closely in the outer regions of the $M_{500} < 10^{13.5}$ halos, however these halos exhibit a growing low luminosity core, with the radial extent of the core increasing with redshift to $\approx 0.1R_{200}$ at $z=0$. $T_X$ profiles show little evolution in shape, however demonstrate a more significant decrease in overall temperature in the lower mass halos, bringing both mass bins in line with each other by $z=0$. $S_X$ profiles flatten with redshift driven by a higher increase in entropy in core regions.   }
    \label{fig:Profiles_fiducial}
\end{figure*}

\begin{figure*}
    \centering
    \includegraphics[width = \textwidth]{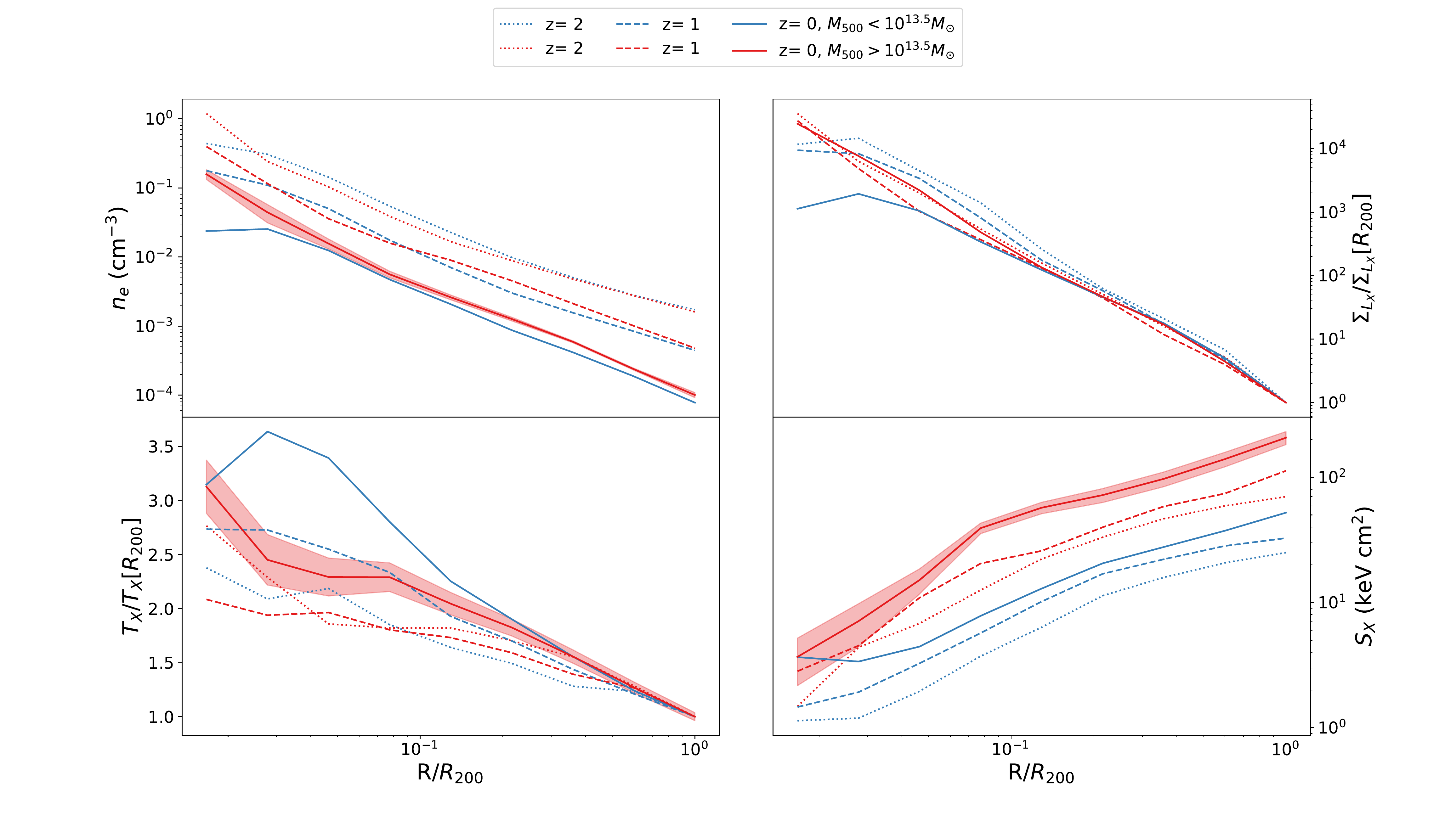}
    \caption{Median profils at $z=2,1,0$ for electron density, $\Sigma_{L_X}$, $T_X$, and $S_X$ in the $50\hmpc$ No-Jet \simba run.  $z=2$ is represented by a dotted line, $z=1$ by a dashed line, and z=0 by a solid line. The median profiles of halos with $M_{500} < 10^{13.5} M_{\odot}$ are coloured blue while those $> 10^{13.5} M_{\odot}$ are represented by the red lines. The $\Sigma_{L_X}$ profiles are scaled to their values at $R_{200}$ to accentuate the evolution in shape however the shift in trends can still be seen in the other plots. $n_e$ profiles experience an overall decrease with redshift, maintaining much steeper profiles than the fiducial runs. There is little difference between mass bins. $\Sigma_{L_X}$ profiles experience very little evolution in shape, with the least massive halos developing a slightly flattened core at $z=0$. $T_X$ profiles experience an overall decrease in both mass bins, however fail to converge at $z=0$.  Little evolution is seen in the shape of the profiles at either mass bin. $S_X$ profiles experience a smaller increase especially at lower masses, with very little evolution seen in the shape of the profiles.}
    \label{fig:Profiles_50_s50nojet}
\end{figure*}

Global $L_X-$weighted measures tend to weight central regions more heavily where much of the X-ray emission arises, resulting in the properties used in these scaling relations being more sensitive to the core.  It is interesting to ask where in radius the various feedback processes in \simba impact the observable properties. To garner a more detailed view of where feedback impacts the halo we examine the evolution of these halos' profiles. Thus we examine the evolution of profiles in the fiducial $100\hmpc$ \simba run for electron density ($n_e$), X-ray surface brightness $\Sigma_{L_X}$, $T_X$, and entropy $S_X$, and subsequently compare with the No-Jet run since we have shown that the jets have the largest impact on the hot gas properties.

We generate projected (2-D) profiles for halos with $10^{12.5}< M_{500} < 10^{13.5} M_{\odot}$ and $M_{500} > 10^{13.5} M_{\odot}$, by stacking halos in these mass bins scaling each to their respective $R_{200}$ values. These profiles are plotted for $z=2,1,0$. Profiles at $z=3$ show similar trends as at $z=2$ and so were omitted for clarity. 
\subsection{ Full \simba Profiles }

Figure~\ref{fig:Profiles_fiducial} shows the median profiles for electron density ($n_e$) in the top left plot, $L_X$ surface density ($\Sigma_{L_X}\equiv L_X/A$ where $A$ is the surface area of the 2-D radial bin) scaled by its value at $R_{200}$ in the top right, $T_X$ scaled by its value at $R_{200}$ in the bottom left, and Entropy ($S_X$) in the bottom right. Here Entropy is defined as $T_X/n_e^{2/3}$ and is calculated by dividing the temperature of each particle by its $n_e^{2/3}$ and then calculating the $L_X$ weighted average across the radial bin. Profiles are represented by dotted, dashed, and solid lines for $z=2, z=1$, and $z=0$ respectively. Profiles of halos with $M_{500} < 10^{13.5} M_{\odot}$ are blue and halos with $M_{500} > 10^{13.5} M_{\odot}$ are red. The shaded areas represent the standard error and are shown on halos with $M_{500} > 10^{13.5} M_{\odot}$ at $z=0$. We scale $\Sigma_{L_X}$ and $T_X$ by their values at $R_{200}$ in order to highlight the evolution in the shapes of these profiles; the amplitude evolution for these quantities was quantified in the previous section.

The top left panel shows the electron density profiles.  The most visible trend in both mass bins is a steady decrease in the amplitude with time.  Less evident but still clear is a slight flattening, indicating a stronger decrease in core density.  At $z=2$ and $z=1$ the profiles for $M_{500} < 10^{13.5} M_{\odot}$ and $M_{500} > 10^{13.5} M_{\odot}$ lie very close to each other, consistent with self-similarity.  However, at $z=0$ we see that the two mass bins have diverged in amplitude, with more massive halos having an overall higher electron density.   Notably, this occurs at all radii, and is not limited to the core.  This is consistent with the stronger removal of hot gas due to jet feedback that heats the gas in lower mass halos, but is not expected in a scenario where more efficient cooling in lower mass systems depletes the hot gas in the central region.

Moving to the temperature profiles in the lower left panel of Figure~\ref{fig:Profiles_fiducial}, we first note that most of the profiles are within a factor of two of being isothermal (note the small range in $y$-axis).  Within this, however, some notable differences in profile shapes are seen.  Even at early epochs, high-mass halos have a steeper profile than low-mass ones.  Both mass bins have steeper profiles with time, and by $z=0$ they exhibit a mildly cooler core at $R\ll 0.1R_{200}$.  None of these core drops are so dramatic as to identify these systems as ``cool core" systems, which have $T_X$ profiles dropping well below $T_X(R_{200})$ in the centres.  While some individual profiles do this, the mean profile is not a classic cool core system in either mass bin.  The variation among individual profiles and their central cooling times is an interesting study that we leave for future work; here we simply note that there are mild differences in profile shapes in the two mass bins over time.

The electron density plus the temperature are the main determinants of the X-ray luminosity.  Figure~\ref{fig:Profiles_fiducial}, top right, shows $\Sigma_{L_X}$ profiles scaled to their value at $R_{200}$.  Overall, the profiles in both mass bins becomes slightly shallower with time, showing a pronounced central peak at $z=2$ reflective of the peak in the $n_e$ profile then.  Already at $z=1$, differences emerge in the core, with a slight flattening of the inner profile between $z=2\to 1$ in the low-mass halos, becoming even more pronounced at $z=0$.  The outer ($R\ga 0.2R_{200}$) profiles remain fairly consistent at all redshifts, indicating that the drop in X-ray luminosity for low-mass systems at late times (and the associated departures from self-similarity) arises primarily from the core region, likely indicating the effect of jets on the cores of lower mass halos. 

Finally we investigate the X-ray weighted entropy profiles in the bottom right plot of Figure~\ref{fig:Profiles_fiducial}. Here the combined effect of $T_X$
and $n_e$ can be seen as given by the $S_X \equiv T_X/n_e^{2/3}$ entropy definition. A steady increase in entropy can be seen in both mass bins, with lower mass halos exhibiting more flattening across redshift. While both mass bins show a similar raising of core entropies, lower mass halos do appear to exhibit overall lower entropies at early times. This is consistent with the idea that cooling is more effective in low-mass halos at early epochs, but by $z=0$, the jet feedback has evacuated gas such that it raises the entropy particularly strongly in low-mass systems.  This raised entropy is likely a critical factor to suppress cooling in these systems in order to keep galaxies quenched.

\subsection{No-Jet Profiles}

We have implicated jets in impacting the evolution of low-mass profiles in comparison with higher-mass systems.  We can test this more directly by examining the same set of profiles in our No-Jet run.  We note that here we must use the $50\hmpc$ run here, but we have checked that the profiles in the $50\hmpc$ full \simba run follows the same trends as those in the $100\hmpc$ fiducial volume that we presented in the previous section.  However, this does mean that the $M_{500} > 10^{13.5} M_{\odot}$ bin is notably less populated in the No-Jet case owing to the smaller volume.

Figure \ref{fig:Profiles_50_s50nojet} shows the profiles in the No-Jet run, 
formatted as in Figure~\ref{fig:Profiles_fiducial}, displaying profiles at $z=0,1,2$ (solid, dotted, dashed lines) in high and low halo mass bins (red and blue, respectively).

For the $n_e$ profiles, in the top left plot, we see two interesting effects: Firstly there is significantly less evolution across mass scales, with only an order of magnitude decrease in density, as opposed to the 3 orders of magnitude decrease seen in Figure~\ref{fig:Profiles_fiducial}. This difference in evolution is most notable in the core regions, with all halos maintain a much higher core density over redshift, even while the electron density at $R_{200}$ has fairly similar evolution with and without jets; the electron density profiles without jets are thus much steeper than with jets.

Secondly the two mass bins evolve in line with each other, with the only difference being a flatter inner core electron density seen in the lower mass halos.  Thus without jets, the profiles are much more in line with self-similarity, with perhaps a mild effect from other feedback processes concentrated at $\la 0.03R_{200}$ in lower-mass groups.  Comparing with Figure~\ref{fig:Profiles_fiducial}, this implicates jets in lowering the electron densities across the entire mass range, and doing so across all radii.  

The $T_X$ profiles in the lower left panel of Figure~\ref{fig:Profiles_50_s50nojet} also experience a similarly mild evolution. The shapes of the profiles remain fairly constant in both mass bins, with halos maintaining higher temperature cores.  In detail, the $T_X$ profiles are actually steeper for low mass halos vs. high-mass ones, which is opposite to the trend seen in the full-physics \simba case.  By $z=0$, the inner profiles in the low-mass groups shows a significant elevation, possibly due to enhanced energy input from other feedback processes combined with cooling out of the lower-temperature gas to form into stars.  Comparing to Figure~\ref{fig:Profiles_fiducial}, the No-Jet profiles are clearly much steeper, and the effect of jets here is already evident at $z=2$ with steeper $T_X$ profiles.  This may indicate that even early low level jet feedback works to push hot X-ray emitting gas from cores to the outer regions.  This trend strengthens down to $z=0$, particularly in the low-mass groups.

In the top right plot of Figure~\ref{fig:Profiles_50_s50nojet}, the $\Sigma_{L_X}$ profiles are again notable for showing very little evolution in the shape of the profiles, except for a slightly lower core luminosity seen in the lower mass halos. This directly mimics the trend seen in the electron densities.  Moreover, by $z=0$, the $\Sigma_{L_X}$ profiles are definitely steeper than in the fiducial run in Figure~\ref{fig:Profiles_fiducial}.  This most clearly demonstrates that it is AGN jet feedback that is primarily responsible for breaking the self-similar scaling of the X-ray emission.  Other feedback and cooling effects have a much smaller impact.

Finally in the bottom right panel of Figure~\ref{fig:Profiles_50_s50nojet} we show the entropy profiles.  We see a steady but slow increase in the normalisation of the entropy profiles. The lower mass halos fail to come in line with the most massive halos, and most notably neither bin experiences a rise of core entropy with respect to the outer regions. This is in line with the idea that jets work to remove low entropy gas, especially from the cores of halos.  The low central entropy allows for substantial cooling that keeps galaxies fueled and star-forming even in high-mass halos in the No-Jet run, which does not agree with observations.  This highlights how jets are responsible for quenching star formation in massive systems, via their impact on the hot gas in the cores of massive halos.

Overall we see little to no evolution in the shape of the X-ray profiles in the No-Jet case, along with a smaller evolution in the normalisation of these profiles.  This indicates that jets not only impact the overall properties, but also importantly how these properties are expressed as a function of radius. Interestingly we see jets having an effect in flattening profiles, or creating larger cores (specifically in the low mass $L_X$ profiles).  Finally while jets have a clear impact in the shapes of profiles in both mass bins, larger effects are seen in the normalisations of these profiles in the $M_{500} < 10^{13.5} M_{\odot}$ bin, highlighting the greater impact of jet feedback on lower mass halos.

\section{Summary}
 In this paper we have examined the redshift evolution of the X-ray scaling relations and X-ray profiles of intragroup gas within the \simba suite of cosmological hydrodynamical galaxy formation simulations.  We consider the scaling relations at particular redshifts from $z=3\to 0$, the median paths of halos in different mass bins through scaling relation space, and the hot gas and X-ray profiles of intragroup gas.  We isolate the physical processes driving this evolution by comparing among feedback variant \simba runs with different feedback models turned on an off, in comparison with the baseline model of self-similar evolution.  We show that: 
 
 \begin{itemize}
     \item \simba halos show are consistent with a self-similar slope of their X-ray scaling relations at $z\ga 1.5$, while at $z\la 1.5$ there are an increasing number of groups that are substantially deviated from self-similarity.
     \item Evolution of the $L_X-M_{500}$ scaling relation sees significant change occurring after $z\la 1.5$ driven primarily by a decrease in $L_X$ among halos with $M_{500} < 10^{13.5} M_{\odot}$. This is further corroborated by median tracks of halos showing that the least massive halos show a 2~dex drop in $L_X$ for a 0.3~dex increase in $M_{500}$ while the most massive halos show a slight increase in $L_X$ for an order of magnitude increase in $M_{500}$ from $z=3\to 0$.
     \item Evolution of the $T_X - M_{500}$ appears to be driven by an overall decrease in $T_X$ for halos with $M_{500}<10^{13.5} M_{\odot}$. Following halo over time indicate that the least massive halos at $z=3$ do not lower in $T_X$ suggesting that halos crossing above $10^{12.3} M_{\odot}$ at later epochs are responsible for the lowering of $T_X$ in the smaller halos.
     \item The hot gas fraction $f_{\rm gas}$ vs. $M_{500}$ shows dramatic evolution at $z\la 1.5$ with a large drop and increasing in scatter particularly for $M_{500} < 10^{13.5} M_{\odot}$ systems. This scatter is highly correlated with the black hole mass fraction $f_{BH}\equiv M_{BH}/M_{500}$, such that halos with central galaxies having high $f_{BH}$ have low $f_{\rm gas}$ and vice-versa.
     Median tracks show that halos with $M_{500} > 10^{13.5} M_{\odot}$ have hot gas fractions encompassing the majority of the cosmic baryon fraction, while $M_{500} < 10^{13.5} M_{\odot}$ halos show an order of magnitude decrease. 
    
     \item The X-ray weighted metallicity shows a significant trend with the stellar baryon fraction in halos at early epochs, which gets washed out over time.  Larger groups have a metallicity around one-third solar, while smaller groups generally have slightly higher metallicity but with a very large scatter.
     
     \item The $L_X-T_X$ relation shows an increasing departure from self-similarity with time, with a growing population at $T_X\la 1$~keV having lower $L_X$ that becomes particularly prominent at $z\la 1.5$. The halos that are most deviant from self-similarity are the ones with the highest $f_{BH}$, circumstantially implicating black hole feedback as the cause of this deviation.
     
     \item To pinpoint the particular physical processes driving the above evolution, we examine smaller $50\hmpc$ \simba runs with specific feedback modules turned off.  Turning off both AGN and star formation feedback results in little evolution in $f_{\rm gas}$, and modest evolution in $L_X$ and $T_X$ mostly consistent with self-similarity.
     
     \item Introducing stellar (but no AGN) feedback does not change $f_{\rm gas}$, $L_X$, or $T_X$, but it does result in significantly earlier deposition of metals into the hot gas even by $z=3$.
     
     \item Introducing AGN jet feedback has a dramatic impact on scaling relations.  Jets are clearly implicated in lowering 
     $f_{\rm gas}$ in $M_{500} < 10^{13.5} M_{\odot}$ groups, particularly at $z\la 1.5$.  This reflects in a dramatic reduction in $L_X-M_{500}$ in such groups over time, and an increased reduction in $T_X-M_{500}$ with time. 
     
     \item Remarkably, the $L_X-T_X$ evolution undergoes a sort of cancellation, in that the lowering of both $L_X$ and $T_X$ results in groups evolving mostly along the relation.  The net effect is that groups today end up much more spread out along the $L_X-T_X$ relation, but the overall relation is surprisingly similar in slope and amplitude in all the feedback variants.
     
     \item The $Z_X - M_{500}$ evolution is unique in that it is more impacted by stellar feedback than AGN feedback. With no feedback, a steady increase is seen in metallicity. The inclusion of stellar feedback produces hot gas metallicities $>0.1Z_\odot$ even by $z=3$. The inclusion of jet and X-ray feedback then seems to reintroduce a modest level of upwards evolution in metallicity particularly in low-mass halos, resulting in the final $z=0$ relation that increases mildly to low halo masses.
     
     \item Electron density profiles in \simba drop and flatten with time, at all masses.  The drop in smaller groups from $z\sim 1\to 0$ is however significantly greater than in larger systems.  
     $L_X$ profiles also experience a flattening at all mass ranges, however it is most prominent at $M_{500} < 10^{13.5}M_{\odot}$, growing a constant surface brightness core ($R\la 0.1R_{200}$) by $z=0$.  The temperature evolution is not dramatic, but entropy is also flattened, resulting in all groups having median core entropies above 100~keV~cm$^2$. 
     \item Turning off jet feedback has significant impact on both  the shape and normalisation of these profiles. Electron density, temperature, and luminosity profiles are steeper.  There is much less difference between the two mass bins, with only a very small core appearing at late times in the lower mass groups.  The entropy profile is much steeper, with typical entropies at $\sim 0.01R_{200}$ below 10~keV~cm$^2$ at all masses. 
    
\end{itemize}
 
These results overall show that the implementation of AGN jet feedback in \simba has a significant impact on X-ray properties both globally and in terms of their profiles.  It is this feedback that causes strong departures from self-similarity, which tend to become noticeable around $z\sim 1.5$ as AGN jets turn on in central group galaxies, and that results in intragroup gas that closely matches X-ray observations at $z=0$ as shown in \citet{Robson:2020}.  The evolution of X-ray profiles provides further constraints on these models, demonstrating the effects of feedback in evacuating low entropy gas and creating X-ray surface brightness cores particularly in lower-mass groups.
     
A dearth of observations at these mass scales at $z\gg 0$, along with an overall lack of consensus caused by selection biases in available results, makes it premature to compare the evolution of \simba's groups to X-ray observations. With future missions such as {\it Lynx} and {\it Athena} providing significantly deeper observations that will push toward higher redshifts, it should be possible to further constrain these models providing even greater insight into exactly how AGN feedback impacts the growth of groups and clusters. In future work we plan to examine how these next-generation telescopes will help uncover the physics of gas within groups and clusters, with the successes of \simba suggesting that it provides a suitable framework for interpreting these observations.

\newpage
\section*{Acknowledgements}

The authors thank Weiguang Cui, Nicole Thomas, Sarah Appleby, and Rohit Kondapally for helpful discussions.
RD acknowledges support from the Wolfson Research Merit Award program of the U.K. Royal Society.
This work used the DiRAC@Durham facility managed by the Institute for Computational Cosmology on behalf of the STFC DiRAC HPC Facility. The equipment was funded by BEIS capital funding via STFC capital grants ST/P002293/1, ST/R002371/1 and ST/S002502/1, Durham University and STFC operations grant ST/R000832/1. DiRAC is part of the National e-Infrastructure.

\section*{Data Availability}
The raw simulation data and catalogs for all the runs used in this paper are available at {\tt https://simba.roe.ac.uk}.  The X-ray data will be made available on request to the lead author.
\label{lastpage}
\bibliographystyle{mnras}
\bibliography{references.bib} 

\end{document}